\def\VersionFinal{}

\ifdefined\VersionLong%
	\newcommand{\LongVersion}[1]{\#1}
	\newcommand{\ShortVersion}[1]{}
\else
	\newcommand{\LongVersion}[1]{}
	\newcommand{\ShortVersion}[1]{#1}
\fi

\documentclass[runningheads]{llncs}

\usepackage{comment}
\ifdefined\VersionLong%
    \includecomment{LongVersionBlock}
    \excludecomment{ShortVersionBlock}
\else
    \excludecomment{LongVersionBlock}
    \includecomment{ShortVersionBlock}
\fi
\ifdefined\VersionFinal%
    \includecomment{FinalVersionBlock}
\else
    \excludecomment{FinalVersionBlock}
\fi

\usepackage[T1]{fontenc}
\usepackage{graphicx}
\usepackage[square,comma,numbers,sort&compress,sectionbib]{natbib}

\usepackage{xspace}
\usepackage{amsmath, amssymb,latexsym,amsfonts,amstext} 
\usepackage{multirow}
\usepackage{algorithm}
\usepackage{algorithmicx}
\usepackage[noend]{algpseudocode}

\usepackage{listings, graphicx}
\usepackage{booktabs}
\usepackage{array}
\usepackage{color}
\usepackage[svgnames,table]{xcolor}
\definecolor{darkblue}{rgb}{0.0,0.0,0.6}
\definecolor{darkgreen}{rgb}{0, 0.5, 0}
\definecolor{darkpurple}{rgb}{0.7, 0, 0.7}
\definecolor{darkblue}{rgb}{0, 0, 0.7}
\usepackage{float}
\usepackage{subcaption}
\usepackage{mathtools}

	\usepackage[%
			pdfauthor={Amit Gurung, Masaki Waga, and Kohei Suenaga},
			pdftitle={Learning nonlinear hybrid automata from input--output time-series data},
			breaklinks  = true,
			colorlinks=true,
			citecolor   = blue!50!blue,
			linkcolor   = darkblue,
			urlcolor    = blue!50!green,
		]{hyperref}

\usepackage[capitalise,english,nameinlink]{cleveref} 
\usepackage{hyperref}
\crefname{line}{\text{line}}{\text{lines}} 
\crefname{item}{\text{item}}{\text{items}} 
\crefname{example}{\text{Example}}{\text{Examples}} 
\crefname{assumption}{\text{Assumption}}{\text{Assumptions}} 
\crefname{algorithm}{\text{Algorithm}}{\text{Algorithms}}

\usepackage{wrapfig}
\usepackage{paralist} 
\usepackage{xspace}

\newenvironment{ienumeration}
	{\ifdefined\VersionLong\begin{enumerate}\else\begin{inparaenum}[\upshape(i)]\fi}
	{\ifdefined\VersionLong\end{enumerate}\else\end{inparaenum}\fi}

\newcommand{\defProblem}[3]
{%
\noindent\fcolorbox{black}{blue!15}{

	\smallskip

	\begin{minipage}{.95\columnwidth}
		\textbf{#1 problem:}\\
		\textsc{Input}: #2\\
		\textsc{Output}: #3
	\end{minipage}
}

	\medskip

}

\usepackage{tikz}
\usetikzlibrary{arrows,automata,positioning,math}
\tikzstyle{every node}=[initial text=]\ifdefined\VersionFinal%
    \includecomment{FinalVersionBlock}
\else
    \excludecomment{FinalVersionBlock}
\fi
\tikzstyle{location}=[rectangle, rounded corners, minimum size=12pt, draw=black, fill=blue!10, inner sep=2pt]
\tikzstyle{invariant}=[draw=black, dotted, inner sep=1pt] 
\tikzstyle{final}=[double]
\tikzstyle{accepting}=[final]

\ifdefined\VersionWithComments%
	\usepackage[colorinlistoftodos,textsize=footnotesize]{todonotes}
\else
	\usepackage[disable]{todonotes}
\fi

\usepackage[pagewise,switch]{lineno} 
\ifdefined\VersionWithComments\linenumbers%
\fi

\def \X {\ensuremath{\mathcal{X}}\xspace}
\def \I {\ensuremath{\mathcal{I}}\xspace}
\def \O {\ensuremath{\mathcal{O}}\xspace}
\def \H {\ensuremath{\mathcal{H}}\xspace}
\def \L {\ensuremath{\mathcal{L}}\xspace}

\def \G {\ensuremath{\mathcal{G}}\xspace}
\def \M {\ensuremath{\mathcal{M}}\xspace}
\def \Trans {\mathit{Trans}}
\def \Inv {\mathit{Inv}}
\def \Init {\mathit{Init}}
\def \Flow {\mathit{Flow}}
\def \seg {\mathit{sg}}

\def \Natural {\ensuremath{\mathcal{N}}\xspace}
\def \Reals {\ensuremath{\mathcal{R}}\xspace}

\newcommand\CORREL{\mathrm{correl}}
\newcommand{\PROJ}[2]{\mathit{pr}_{#1} ({#2})}

\newcommand\POW{\mathcal{P}}

\usepackage{lineno} 

\usepackage{footnote} 

\newcommand{\DTWDist}{\mathrm{DTW}_{\mathrm{dist}}}
\newcommand{\DTWCorrel}{\mathrm{DTW}_{\mathrm{correl}}}
\newcommand{\forwardBDF}{\mathrm{f_F}}
\newcommand{\backwardBDF}{\mathrm{f_B}}
\newcommand{\thresholdDiffForwardBackwardBDF}{\varepsilon_{\mathrm{FwdBwd}}}
\newcommand{\thresholdDiffBackwardBDF}{\varepsilon_{\mathrm{Bwd}}}
\newcommand{\thresholdDistance}{\varepsilon_{\mathrm{dst}}}
\newcommand{\thresholdDiagonality}{\varepsilon_{\mathrm{cor}}}

\newcommand{\BouncingBall}{\textsc{Ball}}
\newcommand{\TwoTankSystem}{\textsc{Tanks}}
\newcommand{\SwitchedOscillator}{\textsc{Osci}}
\newcommand{\ExcitableCells}{\textsc{Cells}}
\newcommand{\EngineTimingSystem}{\textsc{Engine}}

\usepackage{colortbl}
\newcommand{\tbcolor}{\cellcolor{green!25}\bf}

\makeatletter
\def\orcidID#1{\smash{\href{https://orcid.org/#1}{\protect\raisebox{-1.25pt}{\protect\includegraphics{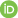}}}}}
\makeatother

\makeatletter
\AtBeginDocument{%
  \@ifpackageloaded{hyperref}
  {\def\@doi#1{\href{https://doi.org/#1}
      {\ttfamily https://doi.org/#1}\egroup}}
  {\def\@doi#1{\ttfamily https://doi.org/#1\egroup}}
  \def\doi{\bgroup\catcode`\_=12\relax\@doi}}
\makeatother

\begin{document}
\title{Learning nonlinear hybrid automata from input--output time-series data}
%
%
\author{Amit Gurung\orcidID{0000-0001-9823-710X} \and
	Masaki Waga\orcidID{0000-0001-9360-7490} \and
	Kohei Suenaga\orcidID{0000-0002-7466-8789}}
\authorrunning{A. Gurung et al.}
%
\institute{Graduate School of Kyoto University, Kyoto, Japan \\
	\email{rajgurung777@gmail.com, \{mwaga, ksuenaga\}@fos.kuis.kyoto-u.ac.jp}}

\maketitle              
%


\begin{abstract}

  Learning an automaton that approximates the behavior of a black-box system is a long-studied problem.
  Besides its theoretical significance, its application to search-based testing and model understanding is recently recognized.
  We present an algorithm to learn a nonlinear hybrid automaton (HA) that approximates a black-box hybrid system (HS) from a set of input--output traces generated by the HS.
  Our method is novel in handling (1) both exogenous and endogenous HS and (2) HA with reset associated with each transition.
  To our knowledge, ours is the first method that achieves both features.
  We applied our algorithm to various benchmarks and confirmed its effectiveness.

\end{abstract}

\section{Introduction}%
\label{section:introduction}

Mathematical modeling of the behavior of a system is one of the main tasks in science and engineering.
If a system exhibits only continuous dynamics, it is well modeled by ordinary differential equations (ODE).
However, many systems exhibit continuous and discrete dynamics, being instances of \emph{hybrid systems (HS)}.
For instance, in modeling an automotive engine, the ODE must be switched following the status of the gear.
A similar combination of continuous and discrete dynamics also appears in many other systems, e.g., biological systems~\cite{DBLP:conf/sfm/BortolussiP08}. 

\begin{wrapfigure}{r}{0pt}
 \begin{tikzpicture}[shorten >=1pt,scale=0.85,every node/.style={transform shape},every initial by arrow/.style={initial text={}}]
  \node[initial,state] (l0) [align=center]{$\dot{x} = v$\\$\dot{v} = -g$};
  %
  \path[->]
  (l0) edge [loop right] node[align=center] {$x \leq 0$/$v \coloneqq -c v$} (l0)
  ;
 \end{tikzpicture}
 \caption{A bouncing ball model}%
 \label{figure:bouncing_ball_example}
\end{wrapfigure}
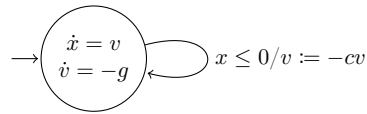

\emph{Hybrid automata (HAs)}~\cite{AlurCHHHNOSY95} is a formalism for HS.\@
\cref{figure:bouncing_ball_example} illustrates an HA modeling a bouncing ball.
In an HA, a set of locations (represented by a circle in \cref{figure:bouncing_ball_example}) and transitions between them (represented by an arrow) expresses its discrete dynamics.
An ODE associated with each location expresses continuous dynamics.
In the HA in \cref{figure:bouncing_ball_example}, the ODEs at the location show the free-fall behavior of the system, and the transition shows the discrete jump caused by bouncing on the floor (i.e., a change in the ball's velocity.)

\begin{table}[tp]
 \newcommand{\CellGood}{\cellcolor{green!20}}
 \newcommand{\CellBad}{\cellcolor{red!20}}
 \centering
 \caption{Comparison of hybrid automata learning methods}%
 \label{table:related_work:comparison}
 \begin{tabular}{l c c c c}
  \toprule
  & Non-linear ODEs?& Exo- and Endogenous? & Infer Resets? & Support Inputs?\\
  \midrule
  \textbf{Ours} & \CellGood{} Polynomial & \CellGood{} Yes{} & \CellGood{} Yes{} & \CellGood{} Yes{}\\
  \cite{10.1145/3556543} & \CellGood{} Polynomial & \CellBad{} Only Exogenous & \CellBad{} No & \CellGood{} Yes \\
  \cite{DBLP:journals/tcps/YangBKJ22} & \CellBad{} Linear & \CellGood{} Yes & \CellBad{} No & \CellGood{} Yes* \\ 
  \cite{DBLP:conf/hybrid/SotoH021, DBLP:conf/atva/SotoHS22} & \CellBad{} Linear & \CellBad{} Only Endogenous & \CellBad{} No & \CellBad{} No\\
  \bottomrule
  \multicolumn{5}{p{1.0\linewidth}}{* \footnotesize{Although this feature is claimed in the paper, the available implementation does not support it.}} 
 
 \end{tabular}

\end{table}

It is a natural research direction to automatically identify an HA given system's behavior.
Not only is it interesting as research, but it is also of a practical impact since learning a model of a black-box system is recently being applied to automated testing (e.g., black-box checking~\cite{DBLP:conf/forte/PeledVY99,DBLP:conf/hybrid/Waga20,DBLP:conf/rv/ShijuboWS21}.)
There have been various techniques to infer an HA from a set of input--output system trajectories.
However, as \cref{table:related_work:comparison} shows, all the existing methods have some limitations in the inferred HA.\@
To the best of our knowledge, there is no existing work that achieves all of the following features: (1) Learned HAs may involve nonlinear ODE as a flow; (2) Learned HAs may be exogenous (i.e., mode changes caused by external events) and endogenous (i.e., mode changes caused by internal events); and (3) Learned HAs may involve resetting of variables at a transition.

This paper proposes an HA-learning algorithm that achieves the three features above.
Namely, our algorithm learns an HA that may be exogenous, endogenous or both.
A learned HA can reset variables at transitions.
These two features make it possible to infer the bouncing ball example in \cref{figure:bouncing_ball_example}, which is not possible in some of the previous work~\cite{10.1145/3556543} despite its simplicity.
Furthermore, an HA learned by our algorithm may involve ODEs with polynomial flow functions, whereas existing work like~\cite{DBLP:journals/tcps/YangBKJ22,DBLP:conf/hybrid/SotoH021,DBLP:conf/atva/SotoHS22} can infer only HAs with linear ODEs.

\begin{figure}[tb]
 \centering
 \includegraphics[width=0.71\linewidth]{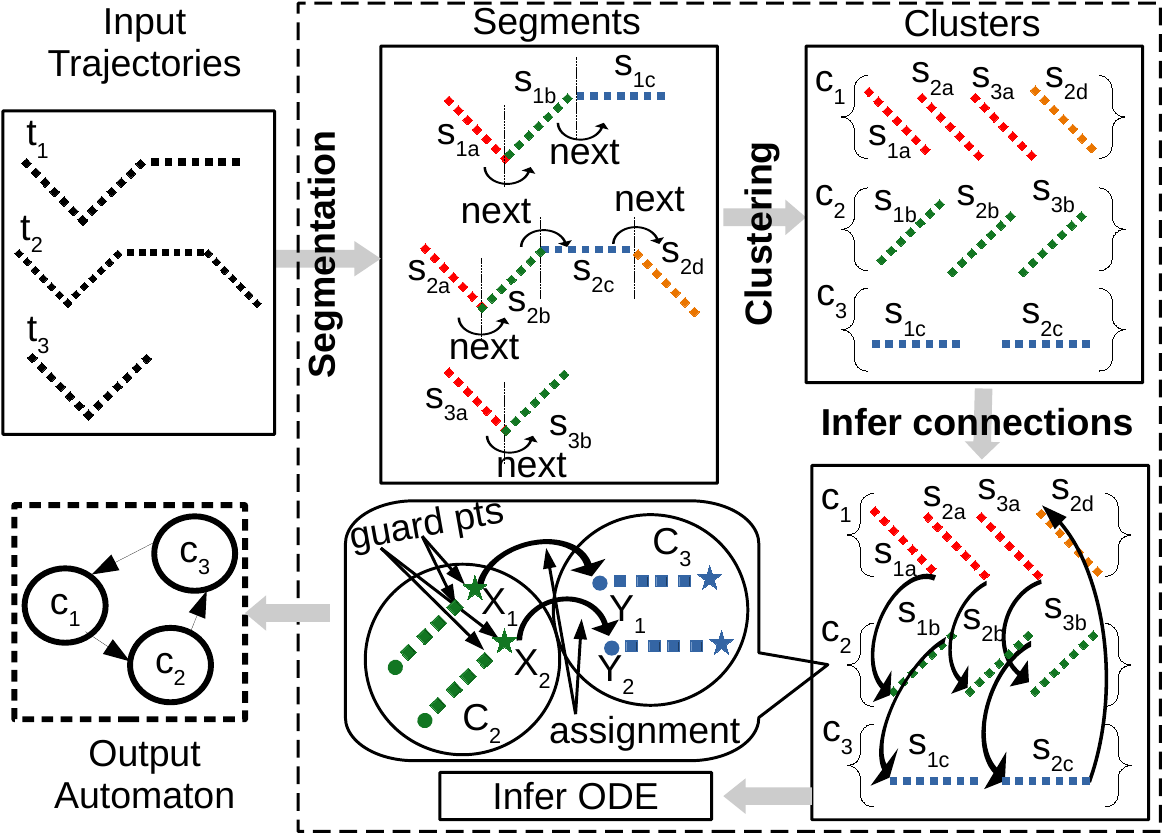}
 \caption{Overview of our HA learning algorithm. In the below center figure, the circle and the star points stand for each segment's first and last points.}%
 \label{figure:outline_of_our_method}
\end{figure}
\cref{figure:outline_of_our_method} shows the overview of our HA learning algorithm.
Our algorithm consists of a \emph{location identification step} and a \emph{transition identification step}.
We explain each step below.

\paragraph{Location identification step.}

To identify the locations, our algorithm first splits trajectories into segments so that each segment consists only of continuous dynamics.
To this end, the algorithm estimates the derivative of each point on a trajectory with the \emph{linear multistep method (LMM)}~\cite{jin2021inferring} and detects the points where the derivative changes discontinuously.

Then, the segments are grouped into clusters so that the segments in each cluster have similar continuous dynamics.
For this, we conduct clustering based on the distance determined by \emph{dynamic time warping (DTW)}~\cite{bellman1959adaptive}, which takes two segments and computes their similarity in terms of the ``shape.''
We treat each cluster as a location of an HA in the following steps.

After the clustering, our algorithm synthesizes an ODE that best describes the continuous dynamics of the segments in each cluster.
Our ODE inference is by a template-based approach.
For each location, we fix a polynomial template---a polynomial whose coefficients are symbols for unknowns---for the flow function of the ODE.
Then, we obtain coefficients of the polynomial via linear regression of the values in a trajectory and the derivative estimated by LMM.

\paragraph{Transition identification step.}

Once locations are identified, our algorithm next synthesizes the transition relation.
It first identifies the pairs of locations (or clusters) between which there is a transition.
Concretely, the algorithm identifies a transition from location $c_1$ to location $c_2$ if (1) there is a segment $s_1$ in $c_1$ and $s_2$ in $c_2$ and (2) $s_1$ immediately precedes $s_2$ in a trajectory.

The algorithm then synthesizes the guard and the reset on each transition.
We synthesize the guard and the reset on each transition in a data-driven manner.
Moreover, we introduce type annotation to improve the inference of resets utilizing domain knowledge;
we explain the method in detail in \cref{inferTransition}.

\paragraph{Contributions}

Our contributions are summarized as follows.
\begin{itemize}
  \item We propose an algorithm inferring a general subclass of HAs from a set of input and output trajectories.
  \item We introduce type annotations to improve the inference of the resets.
  \item We experimentally show that our algorithm infers HAs fairly close to the original system under learning.
\end{itemize}

\paragraph{Related Work}

Despite the maturity of switched-system identification~\cite{GPV2012,jin2021inferring}, only a few algorithms have been proposed to infer HAs.
This scarcity of work in HA learning may be attributed to the additional information that needs to be inferred for HAs (e.g., variable assignments.)

\cref{table:related_work:comparison} summarizes algorithms inferring an HA from a set of trajectories.
In~\cite{DBLP:conf/hybrid/SotoH021, DBLP:conf/atva/SotoHS22}, an HA is learned from a set of trajectories; however, it does not support systems with inputs.
Moreover, only linear ODEs can be learned.
In~\cite{DBLP:journals/tcps/YangBKJ22}, an HA with inputs and outputs is learned from trajectories, but the learned ODEs are still limited to linear functions.
In~\cite{10.1145/3556543}, an HA with polynomial ODEs is learned from inputs and outputs trajectories.
However, the guards in the transitions consist only of the input variables and timing constraints.
Due to this limitation, their method cannot infer an endogenous HA such as the bouncing ball model in \cref{figure:bouncing_ball_example}.
Compared with these methods, our algorithm supports the most general class of HAs, to our knowledge.

We remark that most of the technical ingredients used in our algorithm are already presented in the previous papers.
For example, using LMM for segmentation and inference of polynomial ODEs is also used by Jin et al.~\cite{jin2021inferring} for learning switched dynamical systems and we adapted it for learning HA.
The use of DTW for clustering is common in~\cite{10.1145/3556543}.
We argue that our significant technical contribution is the achievement of learning the general class of HAs by an appropriate adaptation and combination of these techniques, e.g., by projecting the output dimensions during segmentation. 
The use of type annotation to improve the inference of variable assignments is also our novelty, up to our knowledge.



\paragraph{Organization}
After reviewing the preliminaries in \cref{section:preliminaries}, we present our HA learning algorithm and an experimental evaluation of it in \cref{section:methodology} and \cref{section:experiments}, respectively. Finally, we conclude in \cref{conclusion}.

\section{Preliminaries}\label{section:preliminaries}
For a set $X$, we denote its powerset by $\POW(X)$.
For a pair $p := (a,b)$, we write $\PROJ{1}{p}$ for $a$ and $\PROJ{2}{p}$ for $b$.
We denote naturals and reals by $\Natural$ and $\Reals$, respectively.
For vectors $u$, and $v$ with the same dimension,
the \emph{relative difference} between them is
	\begin{math}
		rd(u,v) := \frac{\left\|u-v\right\|}{\left\|u\right\| + \left\|v\right\|}
	\end{math}
	where $\left\|u\right\|$ is the Euclidean norm of $u$. 
        We write $[a, b]$ for the inclusive interval between $a$ and $b$.

\subsection{Trajectories and Hybrid automata}
For a time domain $[0, T] \subseteq \Reals$ and $n \in \Natural$, an \emph{$n$-dimensional (continuous) signal} $\sigma$ is a function assigning an $n$-dimensional vector $\sigma(t) \in \Reals^n$ to each timepoint $t \in [0, T]$.
Execution of a system with $n_1$ dimensional inputs and $n_2$ dimensional outputs can be modeled by an $(n_1+n_2)$-dimensional signal.

A \emph{(discrete) trajectory} is a sequence of vectors with timestamps.
Concretely, an \emph{$n$-dimensional trajectory} $(t_1,x_1), (t_2,x_2), \dots, (t_N,x_N)$ is a finite sequence of pairs of timestamp $t_i \in \Reals$ and the corresponding value $x_i \in \Reals^n$ satisfying $t_1 < t_2 < \dots <t_N$.
For a signal $\sigma\colon [0, T] \to \Reals^n$,
a trajectory $(t_1,x_1), (t_2,x_2), \dots, (t_N,x_N)$ is a \emph{discretization} of $\sigma$
if for any $i \in \{1,2,\dots,N\}$, we have $x_i = \sigma(t_i)$.
We call each vector $(t_i, x_i)$ in a trajectory as a \emph{(sampling) point}.

\emph{Hybrid automata (HAs)}~\cite{AlurCHHHNOSY95,lygeros2008hybrid} is a formalism to model a system exhibiting an interplay between continuous and discrete dynamics.
Since we aim to learn an HA from a set of trajectories with inputs and outputs, we employ HAs with input and output variables.
To define HAs, we fix a finite set of \emph{(continuous state) variables} $\X$, \emph{input variables} $\I$, and \emph{output variables} $\O$ such that $\X = \I \uplus \O$.
A \emph{valuation} is a mapping $\delta \in \Reals^\X$ that represents the value of each variable.

\begin{definition}
 [Hybrid automaton]\label{def:ha}
 A hybrid automaton (HA) $\H$, is a tuple $(\L, \Inv, \Init, \Flow, \Trans)$  where:
	\begin{itemize}
		\item[-] $\L$ is a finite set of \emph{locations};
		
		\item[-] $\Inv: \L \to \POW(\Reals^\X)$ is a function mapping each location $\ell$ to the \emph{invariant} at $\ell$;
		\item[-] $\Init$, the \emph{initial condition}, is a pair $(\ell_0, \delta_0)$ such that $\ell_0 \in \L$ and $\delta_0 \in \Inv(\ell_0)$; 
		\item[-] $\Flow$ is a \emph{flow function} mapping each location $\ell \in \L$ to ODEs of the form $\dot{x} = f(x, u)$, called \emph{flow equation}, where $x$ is the vector of all the variables in $\O$ and $u$ is the vector of all the variables in $\I$; 
		\item[-] $\Trans$ is the set of \emph{discrete transitions} denoted by a tuple $e =(\ell,\G,\M,\ell')$, where $\ell, \ell' \in \L$ are the \emph{source and target locations}, $\G \subseteq \POW(\Reals^{\X})$ is the \emph{guard}, and $\M: \Reals^{\X} \to \Reals^{\O}$ is the \emph{assignment function}. 
	\end{itemize}
 For a transition $e \in \Trans$,
 we write $\G(e)$ and $\M(e)$ for the guard and the assignment function of $e$, respectively. 
\end{definition}

Intuitively, a guard $\G(e)$ of a transition $e$ is the condition that enables the transition: A transition $e$ can be fired if a valuation $\delta$ for the variables satisfies $\delta \in \G(e)$.
An assignment $\M(e)$ specifies how a valuation is updated if the transition $e$ fired: A valuation is updated from $\delta$ to $\delta'$ such that for each $x \in \O$ and $u \in \I$, we have $\delta'(x)= \M(e)(\delta)(x)$ and $\delta'(u) = \delta(u)$ if $e$ is fired.

The semantics of an HA is formalized by the notion of a \emph{run}.
A \textit{state} of an HA $\H$ is a pair $(\ell, \delta)$, where $\ell$ is a location of $\H$ and $\delta \in \Reals^{\X}$ is a valuation. 

\begin{definition}[Run]%
 \label{def_run}
 A \emph{run} of an HA $(\L, \Inv, \Init, \Flow, \Trans)$ is a sequence
 \begin{displaymath}
  (\ell_0,\delta_{0}) \xrightarrow{\tau_0} (\ell_0,\delta_{0}') \xrightarrow{e_0} (\ell_1,\delta_{1}) \xrightarrow{\tau_1} (\ell_1,\delta_{1}') \xrightarrow{e_{1}} \dots \xrightarrow{e_{N-1}} (\ell_N,\delta_{N}) \xrightarrow{\tau_{N}} (\ell_N,\delta_{N}')
 \end{displaymath}
 satisfying $(\ell_0, \delta_{0}) \in \Init$ and for each $i \in \{0,1,\dots,N\}$, there are signals $\sigma^x_i\colon [0, \tau_i] \to \Reals^{\O}$ and $\sigma^u_i\colon [0, \tau_i] \to \Reals^{\I}$ such that 
 \begin{ienumeration}
  \item for any $x \in \O$ and $u \in \I$, we have $\sigma^x_i(0)(x) = \delta_i(x)$ and $\sigma^u_i(0)(u) = \delta_i(u)$, $\sigma^x_i(\tau_i)(x) = \delta_i'(x)$, and $\sigma^x_i(\tau_i)(u) = \delta_i'(u)$,
  \item for any $t \in [0, \tau_i]$, we have $(\sigma^x_i(t), \sigma^u_i(t)) \in \Inv(\ell_i)$ and $\dot{\sigma^x_i}(t) =\Flow(\ell_i)(\sigma^x_i(t), \sigma^u_i(t))$, and
  \item we have $\delta_i' \in \G(e_i)$ and $\delta_{i+1}$ is such that
  for each $x \in \O$ and $u \in \I$, we have $\delta_{i+1}(x) = \M(e_i)(\delta'_{i})(x)$ and $\delta_{i+1}(u) = \delta'_{i}(u)$. 
 \end{ienumeration}
 For such a run $\rho$, a signal $\sigma\colon [0, T_N] \to \Reals^{\X}$ is the signal over $\rho$ if $\sigma$ is such that $\sigma(t)(x) = \sigma^x_i(t - T_{i-1})$ and $\sigma(t)(u) = \sigma^u_i(t - T_{i-1})(u)$ for each $x \in \O$, $u \in \I$, and $i \in \{0,1,\dots,N\}$ such that $T_i \leq t < T_{i+1}$,  where $T_i = \sum_{j \in 0}^{i}\tau_j$ and $T_{N+1} = \infty$.
\end{definition}

\subsection{Linear Multistep Method}\label{sec_LMM}

The \emph{linear multistep method (LMM)}~\cite{butcher2016numerical} is a technique to numerically solve an initial value problem of an ODE $\dot{x(t)} = f(x,t)$.
Concretely, it approximates the value of $x(t_{n+M})$ by using the values of $x(t_{n}), \dots, x(t_{n+M-1})$ and $f(x_{n}, t_{n}), \dots, f(x_{n+M-1}, t_{n+M-1})$---namely, $M$ previous discretized values of $x$ and $f(x,t)$---where $t_{n+i} = t_n + i h$ for some $h > 0$.
For this purpose, LMM assumes the following approximation parameterized over $(\alpha_i)_i$ and $(\beta_i)_i$:
\begin{displaymath}
  \label{eq:lmm:approx}\sum_{i=0}^M \alpha_i x(t_{n-i}) \approx h \sum_{i=0}^M \beta_i f(x(t_{n-i}), t_{n-i}).
\end{displaymath}
Then, LMM determines the values of $(\alpha_i)_i$ and $(\beta_i)_i$ so that the error of the above approximation, quantified with Taylor's theorem, is minimum; see~\cite{butcher2016numerical,suli2003introduction,jin2021inferring} for more detail.
The approximation with the determined values of $(\alpha_i)_i$ and $(\beta_i)_i$ is used to successively determine the values of $x(t)$ from its initial value.

In the context of our work, we estimate the derivative of a trajectory at each point without knowing the ODE.\@
To this end, we use \emph{backwards differentiation formula (BDF)}~\cite{suli2003introduction,keller2021discovery}
 derived from LMM.\@
The idea is to compute the polynomial passing all the points $(t_n, x(t_{n})), \dots, (t_{n+M-1}, x(t_{n+M-1}))$ using Lagrange interpolation~\cite{butcher2016numerical,keller2021discovery} 
 and derive the formula to approximate the derivative at $(t_{n+M}, x(t_{n+M}))$ from the polynomial using LMM.\@ 
Concretely, Lagrange interpolation yields the polynomial: 
$x(t) \approx \sum_{m=0}^M x(t_{n-m}) \prod_{i \ne m}
\frac{t-t_{n-m}}{t_{n-i} - t_{n-m}}$.
By taking the derivative of both sides and setting $t$ to $t_n$, we obtain
$\dot{x}(t_n) = f(x(t_n), t_n) \approx \sum_{m=0}^M x(t_{n-m}) \prod_{i \ne m} (\frac{d}{dt} \left.\frac{t-t_{n-m}}{t_{n-i} - t_{n-m}})\right|_{t=t_n}$.
We use this formula to estimate the derivative at each point in a trajectory.
For instance, the formula to estimate the derivatives with $M = 2$ is: $f(x(t_n)) = \frac{1}{h} ( \frac{3}{2} x(t_n) - \frac{4}{2} x(t_{n-1}) + \frac{1}{2} x(t_{n-2}) )$~\cite{suli2003introduction}.

The above formula estimates the derivative at $x(t_n)$ using $M$ previous points---hence called \emph{backward BDF}.
Dually, we can derive a formula that estimates the derivative at $x(t_n)$ using $M$ following points called \emph{forward BDF}.
We use both in our algorithm.

\subsection{Dynamic Time Warping (DTW)}

Our algorithm introduced in \cref{section:methodology} first splits given trajectories so that each segment includes only continuous dynamics.
Then, it classifies the generated segments based on the ``similarity'' of the ODE behind.
For the classification purpose, we use \emph{dynamic time warping (DTW)}~\cite{bellman1959adaptive}---one of the methods for quantifying the similarity between time-series data in their shapes---as the measure of the similarity inspired by~\cite{10.1145/3556543}.
The previous work~\cite{10.1145/3556543} applies DTW for HA learning and confirms its effectiveness.

The DTW distance between two time-series data $X := (x_1, x_2, \dots , x_M)$ and $Y := (y_1, y_2, \dots , y_N)$, where $M,N \in \Natural$, is defined as follows.
The \emph{alignment path} between $X$ and $Y$ is a finite sequence $P := (p_1, \dots, p_l)$ where $p_i \in \{1,2,\dots,M\}\times\{1,2,\dots,N\}$ and $P$ is an alignment between $\{1,\dots,M\}$ and $\{1,\dots,N\}$.
Concretely, $P$ should satisfy the following conditions: (1) $p_1 = (1,1)$; (2) $p_l = (M,N)$; (3) $(a_{i+1} - a_i, b_{i+1} - b_i)$ is either $(1, 0)$, $(0, 1)$, or $(1,1)$ for any $(p_i, p_{i+1}) = ((a_i, b_i), (a_{i+1}, b_{i+1}))$.
For example, $((1,1), (1, 2), (2, 3), (3, 3), (3,4))$ is an alignment path between $(x_1, \dots, x_3)$ and $(y_1, \dots, y_4)$.

An alignment path $P = (p_1, \dots, p_l)$ between $X := (x_1, x_2, \dots , x_M)$ and $Y := (y_1, y_2, \dots , y_N)$ determines the sum $d_P := \sum_{i=1}^l ||x_{(\PROJ{1}{p_i})}-y_{(\PROJ{2}{p_i})}||$ of the distances between corresponding points in $X$ and $Y$.
Then, the DTW distance $\DTWDist(X,Y)$ between $X$ and $Y$ is defined by $\min_{P}  d_P$, where $P$ moves all the alignment paths between $X$ and $Y$.
There is an efficient algorithm computing $\DTWDist(X,Y)$ in $O(MN)$ based on dynamic programming~\cite{senin2008dynamic}.

For $X$ and $Y$, let $P$ be the alignment that gives the optimal sum of distances between $X$ and $Y$.
We write $\DTWCorrel(X,Y)$ for $\CORREL(P_1, P_2)$, where $P_1 := (\PROJ{1}{p_1}, \dots, \PROJ{1}{p_l})$,  $P_2 := (\PROJ{2}{p_1}, \dots, \PROJ{2}{p_l})$, and $\CORREL(P_1, P_2)$ is the Pearson product-moment correlation coefficients between $P_1$ and $P_2$.
This value becomes larger if $P_1$ and $P_2$ increase evenly.
Thus, the higher this value is, the more $X$ is similar to $Y$.
The effectiveness of this value in classifying segments is also shown in~\cite{10.1145/3556543}.

\section{HA Learning from Input--Output Trajectories}\label{section:methodology}

Our proposed algorithm is an offline and passive approach for learning automata, which involves observing input-output behavior from a given dataset without interacting with the system during learning. Here, we present our HA learning algorithm from given trajectories.
Our problem setting is formalized as follows.

\defProblem{Passive HA learning}%
{trajectories $\{(t^i_1,x^i_1),(t^i_2,x^i_2),\dots,(t^i_{N_i},x^i_{N_i}) \mid i \in \{1,2,\dots,M\}\}$ that are discretizations of signals over runs of an HA $\overline{\H}$}%
{an HA $\H$ approximating $\overline{\H}$}

Our current algorithm learns an HA such that
\begin{ienumeration}
 \item the invariant of each location is $\mathbf{true}$,
 \item each guard is expressed as a polynomial inequality, and
 \item each assignment function is a linear function.
\end{ienumeration}
We assume that 
\begin{ienumeration}
 \item each location of $\H$ has different ODEs and
 \item for each pair $(\ell, \ell')$ of locations of $\H$, there is at most one transition from $\ell$ to $\ell'$.
\end{ienumeration}

\cref{figure:outline_of_our_method} outlines our HA learning algorithm.
We first present the identification of the locations and then present that of the transitions.

\subsection{Identification of Locations}
We identify the locations of an HA by the following three steps:
\begin{ienumeration}
 \item segmentation of the given trajectories,
 \item clustering of the segments, and
 \item inference of ODEs and initial locations.
\end{ienumeration}

\subsubsection{Segmentation of the Trajectories}\label{sec_segmentation}
The first step in our HA learning algorithm is segmentation. Each trajectory is divided into segments so that the dynamics in each segment are jump-free.
We perform segmentation by identifying the change points---the points where the derivative discontinuously changes---along a trajectory. Our approach builds on Jin et al.'s ~\cite{jin2021inferring} method for learning switched dynamical systems, but we adapted and modified it to extend the approach for learning hybrid systems.

\begin{algorithm}[tb]
 \caption{Outline of our segmentation algorithm}%
 \label{algorithm:segmentation}
 \begin{algorithmic}[1] 
  \Require{A trajectory $\tau = (t_1,x_1),(t_2,x_2),\dots,(t_{N},x_{N})$, the step size $M$ in BDF, and the thresholds $\thresholdDiffForwardBackwardBDF$ and $\thresholdDiffBackwardBDF$}
  \Ensure{$\mathit{cp} \subseteq \{1,2,\dots,N\}$ is the set of change points}
  \State{$\mathit{candidates} \gets \emptyset$;\;$C \gets \emptyset$}
  \ForAll{$i \in \{M+1,M+2,\dots,N-M\}$}
    \State{$\mathit{fwd}_i \gets \forwardBDF(\tau|_{\O}, i, M)$} \Comment{Compute the forward BDF}
    \State{$\mathit{bwd}_i \gets \backwardBDF(\tau|_{\O}, i, M)$} \Comment{Compute the backward BDF}
    \If{$rd(\mathit{fwd}_i, \mathit{bwd}_i) > \thresholdDiffForwardBackwardBDF$}
      \State{\textbf{add} $i$ \textbf{to} $\mathit{candidates}$}
    \EndIf{}
  \EndFor{}
  \While{$\mathit{candidates} \neq \emptyset$}
    \State{$i \gets \min(\mathit{candidates})$;\; \textbf{remove} $i$ \textbf{from} $\mathit{candidates}$}
    \If{$i + 1 \not\in \mathit{candidates}$ \textbf{or} $rd(\mathit{bwd}_i, \mathit{bwd}_{i+1}) \geq \thresholdDiffBackwardBDF$}
      \State{\textbf{add} $i$ \textbf{to} $\mathit{cp}$}
      \While{$i + 1 \in \mathit{candidates}$}
        \State{\textbf{remove} $i+1$ \textbf{from} $\mathit{candidates}$; $i \gets i+1$}
      \EndWhile{}
    \EndIf{}
  \EndWhile{}
 \end{algorithmic}%
\end{algorithm}
\begin{figure}[tb]
 \begin{subfigure}[t]{.32\linewidth}
  \centering
  \includegraphics[width=.60\linewidth]{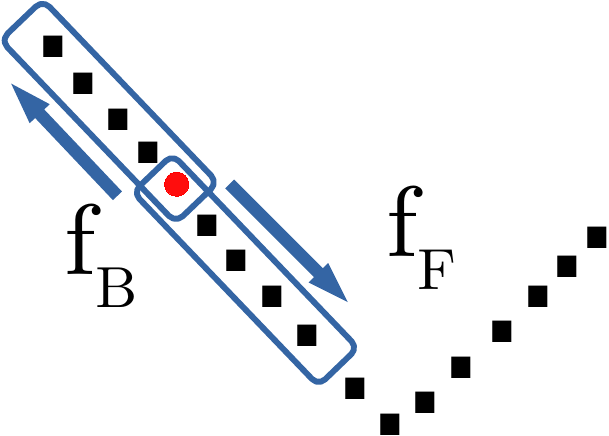}
  \caption{$\mathit{fwd}_i \approx \mathit{bwd}_i$}%
  \label{figure:illustration_segments:not_candidate}
 \end{subfigure}
 \hfill
 \begin{subfigure}[t]{.32\linewidth}
  \centering
  \includegraphics[width=.60\linewidth,page=2]{./figures/segmentation2-crop.pdf}
  \caption{$rd(\mathit{fwd}_i, \mathit{bwd}_i) \gg 0$ and $\mathit{bwd}_i \approx \mathit{bwd}_{i+1}$}%
  \label{figure:illustration_segments:candidate_but_not_change}
 \end{subfigure}
 \hfill
 \begin{subfigure}[t]{.32\linewidth}
  \centering
  \includegraphics[width=.60\linewidth,page=1]{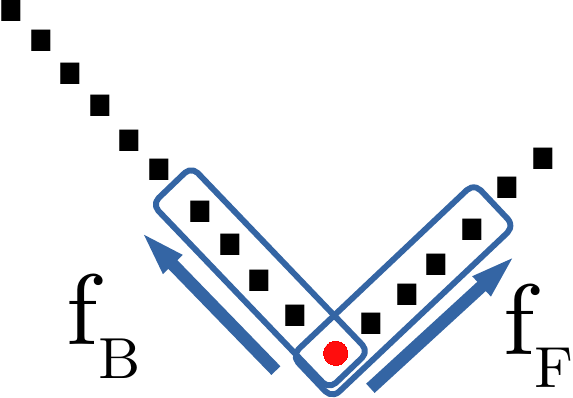}
  \caption{$rd(\mathit{fwd}_i, \mathit{bwd}_i) \gg 0$ and $rd(\mathit{bwd}_i, \mathit{bwd}_{i+1}) \gg 0$}%
  \label{figure:illustration_segments:change}
 \end{subfigure}
 \caption{Illustration of our segmentation algorithm near a boundary of a segment. The red circle in the right figure is the change point because it is the first point satisfying $rd(\mathit{fwd}_i, \mathit{bwd}_i) > \thresholdDiffForwardBackwardBDF$ and $rd(\mathit{bwd}_i, \mathit{bwd}_{i+1}) > \thresholdDiffBackwardBDF$.}
 \label{figure:illustration_segments}
\end{figure}

\cref{algorithm:segmentation} outlines our segmentation algorithm.
For simplicity, we present an algorithm for a single trajectory; this algorithm is applied to each trajectory obtained from the system.
First, for each point in the trajectory, we estimate the derivative using forward and backward BDF ($\mathit{fwd}_i$ and $\mathit{bwd}_i$, respectively) and deem the point as a candidate of change points if $rd(\mathit{fwd}_i, \mathit{bwd}_i)$ exceeds the threshold. 
For example, among the three red circles in \cref{figure:illustration_segments}, we have $\mathit{fwd}_i \approx \mathit{bwd}_i$ for the one in \cref{figure:illustration_segments:not_candidate} and $rd(\mathit{fwd}_i, \mathit{bwd}_i) \gg 0$ for the others. Thus, the red circles in \cref{figure:illustration_segments:candidate_but_not_change,figure:illustration_segments:change} are the candidates of the change point.
We remark that $\mathit{fwd}_i$ and $\mathit{bwd}_i$ are computed with the trajectory $\tau|_{\O}$ projected to the output variables, and our segmentation is not sensitive to the change in the input variables $\I$.

When there are consecutive candidates of the change points, we take the first one satisfying $rd(\mathit{bwd}_i, \mathit{bwd}_{i+1}) \geq \thresholdDiffBackwardBDF$ to precisely estimate the change point. Such an optimization is justified under the assumption that there are at least $2M-1$ points between two consecutive mode changes.
For example, in the example shown in \cref{figure:illustration_segments}, the red circle in \cref{figure:illustration_segments:change} is deemed to be the change point because this is the first candidate satisfying $rd(\mathit{bwd}_i, \mathit{bwd}_{i+1}) \geq \thresholdDiffBackwardBDF$.
Note that, in inferring ODEs, we consider only the points within a segment that satisfy $rd(\mathit{fwd}_i, \mathit{bwd}_i) \le \thresholdDiffForwardBackwardBDF$. This means that candidate points near the boundary of change points are excluded if they do not meet this condition. This is similar to the approach proposed by Jin et al.~\cite{jin2021inferring}

The algorithm splits the trajectories at the identified change points into segments; the change points are not included in the segments.
Our approach improves upon Jin et al.~\cite{jin2021inferring} by adapting their approach for learning switched dynamical systems to the problem of learning hybrid systems. Specifically, we identify change points in a twofold manner. While their approach considers all candidates as change points and drops them from resulting segments, we go a step further and determine the closest change point that precisely separates modes. To achieve this, we search for candidate points until the condition $\mathit{bwd}_i \approx \mathit{bwd}_{i+1}$ is no longer satisfied. This adaptation enables us to include candidate points in the segment actively involved in the transition action, leading to more accurate identification of the transition process.

\subsubsection{Clustering of the Segments}\label{clusteringSection}
Then, we cluster the segmented trajectories so that the segments with similar continuous behaviors are included in the same cluster.
For instance, in \cref{figure:outline_of_our_method}, the continuous behaviors in $S_{1a}$, $S_{2a}$, $S_{2d}$, and $S_{3a}$ are similar and hence included in a single cluster.
We use the identified clusters as the set of locations in the resulting HA.\@
This construction is justified when each location has a different ODE.

\begin{algorithm}[tbp]
 \caption{Outline of the clustering of the segmented trajectories}\label{clusterDTW}
 \begin{algorithmic}[1] 
  \Require{Set $\mathit{Sg}$ of segments and thresholds $\thresholdDistance$ and $\thresholdDiagonality$ for distance and diagonality}
  \Ensure{$C = \{C_1, C_2,\dots,C_n\}$ is a set of set of segments such that each $C_i$ is a cluster}
  \State{$C \gets \emptyset$}
  \While{$\mathit{Sg} \neq \emptyset$}\label{outerLoop}
    \State{\textbf{pick} $\seg$ \textbf{from} $\mathit{Sg}$}\label{newClusterCreation}\Comment{We still have $\seg \in \mathit{Sg}$ after picking it.}
    \State{$\mathit{C}' \gets \{\seg' \in \mathit{Sg} \mid \DTWDist(\seg|_{\O}, \seg'|_{\O}) < \thresholdDistance \land \DTWCorrel(\seg|_{\O}, \seg'|_{\O}) > \thresholdDiagonality\}$}\label{mergeSegs}
    \State{$\mathit{Sg} \gets \mathit{Sg} \setminus \mathit{C}'$}\Comment{$\mathit{C}'$ always includes $\seg$, and $\seg$ is removed from $\mathit{Sg}$.}
    \State{\textbf{add} $\mathit{C}'$ \textbf{to} $C$}
  \EndWhile{}
 \end{algorithmic}
  
\end{algorithm}

\cref{clusterDTW} outlines our clustering algorithm. 
The overall idea is, the algorithm picks one segment (\cref{newClusterCreation}) and creates a cluster by merging similar segments (\cref{mergeSegs}).
We use both $\DTWDist$ and $\DTWCorrel$ to determine the similarity between segments. 
We remark that we compare the segments $\seg|_{\O}$ and $\seg'|_{\O}$ projected to the output variables to ignore the similarity in the input variables.

\subsubsection{Inference of ODEs and Initial Locations}\label{inferODE}

Our ODE inference is by a template-based linear regression.
First, we fix a template $\Phi(x; \theta) = \theta_1 f_1(x) + \theta_2 f_2(x) + \cdots + \theta_N f_N(x)$ of the ODE.\@
In our current implementation, each $f_i$ is a monomial whose degree is less than a value specified by a user, but an arbitrary template can be used.
Then, for each cluster $C_i$ and for each output variable $o \in \O$, we construct the set $P_{i,o}$ of points in $C_i$,\footnote{Notice that, as mentioned above, we only consider the points within a segment that satisfy $rd(\mathit{fwd}_i, \mathit{bwd}_i) \le \thresholdDiffForwardBackwardBDF$.} and the derivative of $o$ at this point. Formally, $P_{i,o} = \{(x, \dot{x}(o)) \mid \exists \seg \in C_i.\, x\in \seg\}$. The derivative $\dot{x}(o)$ is, for example, computed by BDF.\@
Moreover, we can reuse the derivative used in \cref{algorithm:segmentation}.
Finally, we use linear regression to compute the coefficients $\theta$ such that for each $(x, \dot{x}(o))\in P_{i,o}$, we have $\dot{x}(o) \approx \Phi(x; \theta)$.

In the resulting HA, the initial locations are the locations such that the corresponding cluster contains the first segment for some trajectories.
Therefore, we have multiple initial locations if there are trajectories such that their first segments do not satisfy the similarity condition during clustering.

\subsection{Identification of Transitions}\label{inferTransition}

\begin{wrapfigure}[10]{r}{0pt}
 \includegraphics[width=.25\linewidth,page=2]{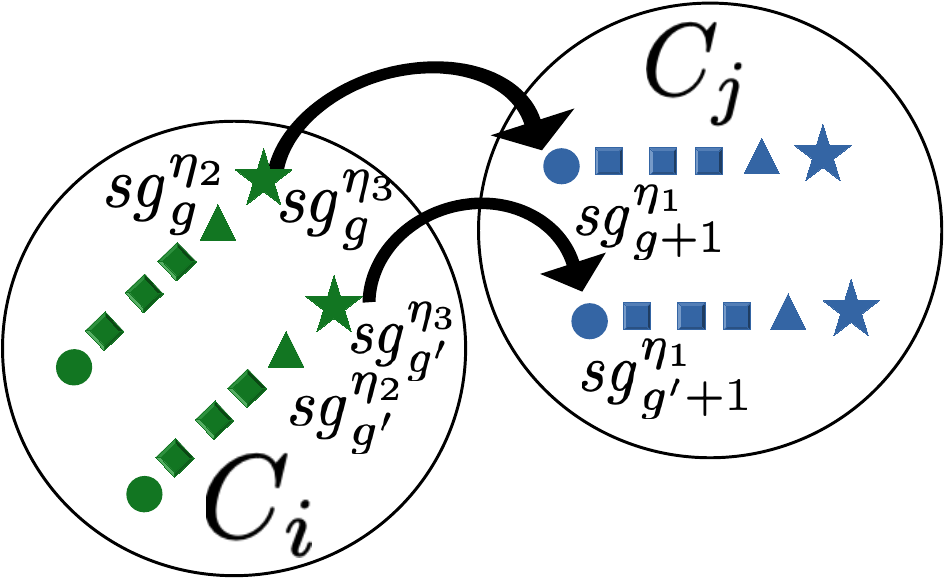}
 \caption{Illustration of the points connecting clusters $C_i$ and $C_j$}%
 \label{figure:illustration_connecting_points}
\end{wrapfigure}
After identifying the locations of the resulting HA by clustering the segments, we construct transitions.
Let $\seg_1, \seg_2, \dots, \seg_m$ be the segments obtained from a single trajectory by \cref{algorithm:segmentation} and ordered in chronological order; segment $\seg_i$ immediately precedes $\seg_{i+1}$ in the original trajectory.
For each segment $\seg_g$, we denote its initial point, the second last point, and the last point by $\seg^{\eta_1}_g$, $\seg^{\eta_2}_g$, and $\seg^{\eta_3}_g$, respectively.

The idea of the transition identification is to make one transition for each triple $(\seg^{\eta_2}_g, \seg^{\eta_3}_g, \seg^{\eta_1}_{g+1})$---called a \emph{connection triple}---and use these points in a triple to infer its guard and assignment; see \cref{figure:illustration_connecting_points} for an illustration.
We note that such a triple is always defined since each segment has at least three points.

Formally, for clusters $C_i$ and $C_j$ and a segmented trajectory $\seg_1, \seg_2, \dots, \seg_m$, the set $T_{i,j}$ of connection triples from $C_i$ to $C_j$ is as follows:
\[
 T_{i,j} = \{(\seg_g^{\eta_2}, \seg_g^{\eta_3}, \seg_{g+1}^{\eta_1}) \mid g \in \{1,2,\dots,m-1\}, \seg_g \in C_i, \seg_{g+1} \in C_j\}
\]
If there are multiple trajectories in HA learning, we construct $T_{i,j}$ for each trajectory and take their union. 

We infer guards and assignments using $T_{i,j}$. 
For each cluster pair $(C_i, C_j)$, the guard of the transition from $C_i$ to $C_j$ is obtained using a support vector machine (SVM) to classify the second last points and the last points.
More precisely, for $T^{\bot}_{i,j} = \{\seg^{\eta_2}_g \mid \exists (\seg_g^{\eta_2}, \seg_g^{\eta_3}, \seg_{g+1}^{\eta_1}) \in T_{i,j}\}$ and 
$T^{\top}_{i,j} = \{\seg^{\eta_3}_g \mid \exists (\seg_g^{\eta_2}, \seg_g^{\eta_3}, \seg_{g+1}^{\eta_1}) \in T_{i,j}\}$,
we compute an equation of hyperplane separating $T^{\bot}_{i,j}$ and $T^{\top}_{i,j}$ using SVM and construct an inequality constraint $\G$ that is satisfied by the points in $T^{\top}_{i,j}$ but not by that in $T^{\bot}_{i,j}$.

For each cluster pair $(C_i, C_j)$, the assignment in the transition from $C_i$ to $C_j$ is obtained using linear regression to approximate the relationship between the valuation before and after the transition.
More precisely, we use linear regression to compute an equation $\M$ such that for each $(\seg^{\eta_2}_g, \seg^{\eta_3}_g, \seg^{\eta_1}_{g+1}) \in T_{i,j}$ and for each $x \in \O$, $\seg^{\eta_1}_{g+1}(x)$ is close to $\M(\seg^{\eta_3}_{g})(x)$. Such $\M$ is used as the assignment.

\subsubsection{Improving Assignments Inference with Type Annotation}

If we have no prior knowledge of the system under learning, we infer assignments using linear regression, as mentioned above.
However, even if the exact system dynamics are unknown, we often know how each variable behaves at jumps.
For instance, it is reasonable to believe that a variable representing temperature is continuous; hence, it does not change its value at jumps.
Such domain knowledge is helpful in inferring precise assignments rather than one using linear regression.

To easily enforce the constraints from domain knowledge on variables, we extend our assignment inference to allow users to annotate each variable with types expressing how a variable is assumed to behave at jumps.
We currently support the following types.
\paragraph{No assignments}
If a variable is continuous at a jump (e.g., the variable representing temperature mentioned above), one annotates the variable with ``no assignments''.
For a variable $x$ with this annotation, the procedure above infers an assignment that does not change the value of $x$.

\paragraph{Constant assignments}

If the value assigned to a variable at a jump is a fixed constant, we annotate the variable with ``constant assignments''.
For instance, in the bouncing ball HA depicted in \cref{figure:bouncing_ball_example}, the variable $x$ is reset to 0 upon reaching the ground or when the guard condition is satisfied. 

\paragraph{Constant pool}
If the value assigned to a variable at a jump is chosen from a finite set, one annotates the variable with ``Constant pool'' accompanied with the finite set $\{v_1,\dots,v_n\}$.
An example of such a variable is one representing the gear in a model of an automotive.
For a variable with this annotation, our algorithm infers the assignment at a jump by majority poll: For a transition from cluster $C_i$ to $C_j$, it chooses the value most frequently occurring in $T_{i,j}$ as $\seg_{g+1}^{\eta_1}$.

\subsection{Impact of Parameter Selection on Model Accuracy} \label{parameter_selection}
We recognize the complexity and potential challenges inherent in ensuring that our proposed hybrid automaton closely emulates the original black-box system, given the intricate and often nonlinear dynamics at play. Nevertheless, our methodology is developed to capture the crucial behaviors of the original system, serving as a foundation for further analysis and understanding.
In our method, specific tuning parameters play pivotal roles during the segmentation and clustering processes. To illustrate, the parameter $\thresholdDiffForwardBackwardBDF$ contributes significantly to effective segmentation, while $\thresholdDiffBackwardBDF$ facilitates pinpointing the correct transition point, which in turn allows for the accurate inference of data points for guards and assignments. It is critical to mention that the choice of these thresholds must be well-thought-out. For instance, a large value for $\thresholdDiffForwardBackwardBDF$ might overlook some change points, while a smaller value could lead to unnecessary redundancy. Therefore, depending on the dynamical class of the system under study, these thresholds may require careful adjustments.

Similarly, during the clustering process, the thresholds $\DTWDist$ and $\DTWCorrel$ play instrumental roles in establishing efficient similarity comparisons between segments. While a $\DTWCorrel$ of 0.8 typically works well in deciding segment similarity, the co-adjustment of $\DTWDist$ and $\DTWCorrel$ parameters effectively allows the clustering process to manage the number of modes in the learned hybrid automaton, balancing precision in the process. In essence, the judicious selection of these parameters is integral to the success of our method in learning an accurate hybrid automaton.

In addition, it is crucial to emphasize that the accuracy of the learned hybrid automaton heavily depends on the amount of data available. We have adapted our segmentation process and ODE inference from the approach presented by Jin et al.~\cite{jin2021inferring}, which provides bounds for estimation errors based on sampling time step and a priori knowledge of system dynamics. While perfect replication of the black-box system may not be achievable, the goal is to construct a meaningful and practical model within the framework of a hybrid automaton. See \cref{table:related_work:comparison} for a comparison of these methods.

\section{Experiments}\label{section:experiments}
We implemented our proposed algorithm using a combination of C++, Python, and MATLAB/Simulink/Stateflow:
The HA learning algorithm is written in Python;
The learned model is translated into a Simulink/Stateflow model by a C++ program;
We use MATLAB to simulate the learned model.
We optimized the ODE inference by using only a part of the trajectories when they were sufficiently many.
We take $M=5$ as the step size for BDF. 
Our implementation is available at \url{https://github.com/rajgurung777/HybridLearner} and the artifact at \url{https://doi.org/10.5281/zenodo.7934743}.

We conducted experiments
\begin{ienumeration}
	\item to compare the performance of our algorithm against a state-of-the-art method and
	\item to evaluate how the type annotation helps our learning algorithm.
\end{ienumeration}
For the former evaluation, we compared our algorithm against one of the latest HA learning methods called POSEHAD~\cite{10.1145/3556543}.
Among the recent hybrid-automata learning methodologies, POSEHAD is the closest to ours in that (1) it handles hybrid systems with nonlinear ODEs and (2) it supports input signals to a system; see \cref{table:related_work:comparison}.
We compared our algorithm with and without a type annotation for the latter evaluation, denoted as ``Type'' and ``W/o Type,'' respectively. 
We also compared our method with two other methods (HAutLearn~\cite{DBLP:journals/tcps/YangBKJ22} and HySynthParametric~\cite{DBLP:conf/atva/SotoHS22}); the result is presented in \cref{sec:comparisonWithOtherMethods}.

In the upcoming comparisons in \cref{result_compare_posehad,sec:comparisonWithOtherMethods}, our proposed method is contrasted with several existing approaches, including POSEHAD, HAutLearn, and HySynthParametric. While noteworthy in their respective areas, these methods have certain limitations when it comes to handling the comprehensive features of hybrid automata - a challenge our method is designed to overcome. Specifically, our approach manages all crucial aspects of hybrid automata, such as guards, nonlinear ODEs, assignments, and support for input signals. It's important to underline that the purpose of these comparisons is not to downplay these existing methods but to highlight the unique scope of functionality and adaptability our method brings to exploring hybrid automata. While this may give an initial impression of an unbalanced comparison, it's essential to emphasize the comprehensive capabilities of our approach, designed to address the gaps left by these other notable methods.

Each benchmark consists of a Simulink/Stateflow model, which we call an \emph{original model}, and two sets of trajectories generated from the original model, which we call \emph{training} and \emph{test} sets.
We generated trajectories by feeding random input trajectories and random initial values of the state variables to the original model.
The training set is used to learn an HA, which we call a \emph{learned model}, and the test set is used to evaluate the accuracy of the learned model.
For each benchmark, the size of the training and test sets are 64 and 32, respectively.

To evaluate the accuracy of the learned model, we feed the same input trajectories and the same initial values to the original and the learned models and compared their output trajectories.
The comparison is based on the DTW distance $\DTWDist$.
A low DTW distance indicates higher accuracy of the learned model. We denote as $\delta_{O_1}$ and $\delta_{O_2}$ the DTW distances between trajectories generated from the original and the learned model on the output variable, $O_1$, and $O_2$, respectively.
We note that, in POSEHAD, the DTW distance is not computed with the entire trajectories but with the segmented trajectories. All the experiments reported in this paper are conducted on a machine with an Intel Core i9 CPU, 2.40GHz, and 32 GiB RAM.\@
We used $\thresholdDiffBackwardBDF = 0.01$ in all our experiments.

\subsection{Benchmark Description}

We briefly describe the benchmarks used in our experiments.

\subsubsection{\BouncingBall} 
This is a benchmark modeling a bouncing ball taken from the demo example of Simulink~\cite{MathWorksBouncingBall}.  \cref{figure:bouncing_ball_example} shows the HA.\@ The acceleration due to the gravity $g$ is taken as input. The range of $g$ is $[-9.9, -9.5]$. We modify the original Simulink model to parameterize the initial values of $x$ and $v$. We also set the model to operate on a fixed-step solver. We let $x \in [10.2, 10.5]$ and $v=15$. The reset factor $c$ in \cref{figure:bouncing_ball_example} is $c = -0.8$.  
We execute the model for a time horizon of 13 units with a sampling time of 0.001, i.e., each trajectory consists of 13,000 points. 
We use $\thresholdDiffForwardBackwardBDF = 0.1$, $\thresholdDistance=9.0$, and $\thresholdDiagonality=0.8$.

\subsubsection{\TwoTankSystem} 
This benchmark models a two tanks system~\cite{hiskens2001stability}.
\cref{fig:twotanks} shows the HA.\@
The system consists of two tanks with liquid levels $x_1$ and $x_2$.
The first tank has in/out flow controlled by a valve $v_1$, 
whereas, the second tank has outflow controlled by the other valve $v_2$.
Both tanks have external in/out flow controlled by the input signal $u$.
There is also a flow from the first tank to the second tank.
In summary, the system has four locations for on and off of $v_1$ and $v_2$. 
The range of the input is $u \in [-0.1, 0.1]$, the initial liquid level of the two tanks are $x_1 =1.2$ and $x_2 = 1$, and the initial location is \textit{off\_off}.
We execute the model for a time horizon of 9.3 units with a sampling time of 0.001, i.e., each trajectory consists of 9,300 points.
We use $\thresholdDiffForwardBackwardBDF = 0.01$, $\thresholdDistance = 1.5$ and $\thresholdDiagonality = 0.7$.

\begin{figure}[tbp]
	\begin{subfigure}[t]{.5\linewidth}
		\centering
		\includegraphics[width=0.75\linewidth]{./figures/Oscillator}
		\caption{An HA model of \textsc{Osci}}%
		\label{fig:oscillator}
	\end{subfigure}
	\hfill
	\begin{subfigure}[t]{.5\linewidth}
		\centering
		\includegraphics[width=0.9\linewidth]{./figures/excitableCellsVariant}
		\caption{An HA model of \textsc{Cells}}%
		\label{fig:excitablecells}
	\end{subfigure}
	\caption{HA models for \textsc{Osci} and \textsc{Cells} benchmarks}%
	\label{figure:excitableCellsModels}
\end{figure}

\subsubsection{\SwitchedOscillator} 
This is a benchmark modeling a switched oscillator without filters~\cite{FLGDCRLRGDM11}.
\SwitchedOscillator{} is an affine system with two variables, $x$ and $y$ oscillating between two equilibria to maintain a stable oscillation.
The HA is shown in \cref{fig:oscillator}.
All the transitions have constant assignments.
This system has no inputs.
The initial values are $x, y \in [0.01, 0.09]$, and the initial location is $\mathit{loc}_1$. We execute the model for a time horizon of 10 units with a sampling time of 0.01, i.e., each trajectory consists of 1,000 points.
We use $\thresholdDiffForwardBackwardBDF = 0.1$,  $\thresholdDistance = 1.0$ and $\thresholdDiagonality = 0.89$.

\subsubsection{\ExcitableCells} 
This is a benchmark modeling excitable cells~\cite{DBLP:conf/hybrid/GrosuMYERS07,ye2005efficient}, which is a biological system exhibiting hybrid behavior. 
We use a variant of the excitable cell used in~\cite{DBLP:conf/cav/SotoHSZ19}. Our HA model is shown in \cref{fig:excitablecells}. 
This model has no inputs.
We take the initial values for the voltage $x \in [-76, -74]$. The \textit{Upstroke} is the initial location. 
We execute the model for a time horizon of 500 units with a sampling time of 0.01, i.e., each trajectory consists of 50,000 points.
We use $\thresholdDiffForwardBackwardBDF = 0.01$, $\thresholdDistance = 1.0$, and $\thresholdDiagonality = 0.92$.

\subsubsection{\EngineTimingSystem} 
This benchmark models an engine timing system taken from the demo examples in the Simulink automotive category~\cite{MathWorksEngineTiming}.
The model is a complex nonlinear system with two inputs and one output signal.
The inputs are the desired speed of the system and the load torque, 
while the output signal is the engine's speed. We simulate the model for a time horizon of 10 units with a sampling time of 0.01, i.e., each trajectory consists of 1,000 points. 
We use $\thresholdDiffForwardBackwardBDF=0.99$, $\thresholdDistance=560$ and $\thresholdDiagonality=0.89$.

\subsection{Results and Discussion} \label{result_compare_posehad}

\subsubsection{Overall Discussion}
\begin{table*}[tb]
	\caption{Summary of the results. The columns $\delta_{O_1}$ and $\delta_{O_2}$ show the minimum (Min), maximum(Max), average (Avg), and standard deviation (Std) of the DTW distance between trajectories generated by the original model and the learned model feeding the test set. The columns Time show the total running time in seconds for learning an HA. Cells with the best results are highlighted.}\label{tab:result_compare}
	\centering

	\begin{tabular}{|c|p{1.23cm}|p{1.1cm}|c|c|p{1.1cm}|c|c|p{1.2cm}|c|c|}
		\hline
		\multirow{2}{*}{Model} & \multirow{2}{*}{Measure} & \multicolumn{3}{c|}{W/o Type} &  \multicolumn{3}{c|}{Type} & \multicolumn{3}{c|}{POSEHAD}\\
		\cline{3-11}
	 &  & $\delta_{O_1}$ & $\delta_{O_2}$ & Time & $\delta_{O_1}$ & $\delta_{O_2}$ & Time & $\delta_{O_1}$ & $\delta_{O_2}$ & Time  \\
		\hline   
	  	\multirow{4}{*}{\textsc{Ball}} & Min($\delta$) & \tbcolor{} 0.15 & \tbcolor{} 0.27 & \multirow{4}{*}{351.2} & \tbcolor{} 0.15 & \tbcolor{} 0.27 & \multirow{4}{*}{336.7} & 127.0 & 1.5e+6 &  \multirow{4}{*}{41535.9} \\			   
	 	\cline{2-4}  \cline{6-7}  \cline{9-10}
	
  	 & Max ($\delta$) & \tbcolor{} 16.4 & \tbcolor{} 12.1 & & \tbcolor{} 16.4 & \tbcolor{} 12.1 &  & 39660.7 & 8.3e+8 & \\
		\cline{2-4}  \cline{6-7}  \cline{9-10}
		
	 & Avg ($\delta$) & \tbcolor{} 1.8 & \tbcolor{} 2.1 & & \tbcolor{} 1.8 & \tbcolor{} 2.1 &  & 9566.2 & 1.7e+8 & \\
		\cline{2-4}  \cline{6-7}  \cline{9-10}
	 & Std ($\delta$) & \tbcolor{} 3.3 & \tbcolor{} 3.1 & & \tbcolor{} 3.3 & \tbcolor{} 3.1  &  & 12695.3 & 1.8e+8 & \\			
		\hline 			
		
		\multirow{4}{*}{\textsc{Tanks}} & Min($\delta$) & 0.1 & 0.3 & \multirow{4}{*}{383.1} & \tbcolor{} 0.003 & \tbcolor{} 0.005 & \multirow{4}{*}{383.5} & 37.8 & 7.8e+12 &  \multirow{4}{*}{13771.5} \\			   
		\cline{2-4}  \cline{6-7}  \cline{9-10}
		
		& Max ($\delta$) & 3.9 &  \tbcolor{}  2.6  &  & \tbcolor{} 3.2 &  \tbcolor{}  2.6 & & 2.3e+4 & 2.0e+14 & \\
		\cline{2-4}  \cline{6-7}  \cline{9-10}
		
		& Avg ($\delta$) & 0.9 & 1.3 &  & \tbcolor{} 0.2 & \tbcolor{} 0.3 & & 8.1e+3 & 9.5e+13 & \\
		\cline{2-4}  \cline{6-7}  \cline{9-10}
		& Std ($\delta$) & 1.0 & 0.68 &  & \tbcolor{} 0.8 & \tbcolor{} 0.74 & & 7.6e+3 & 5.9e+13 & \\			
		\hline

		\multirow{4}{*}{\textsc{Osci}} & Min($\delta$) & 0.21 & 0.3 & \multirow{4}{*}{24.1} & \tbcolor{} 0.17 & \tbcolor{} 0.2 & \multirow{4}{*}{24.9} & 15.8 & 8.8 &  \multirow{4}{*}{404.2} \\			   
		\cline{2-4}  \cline{6-7}  \cline{9-10}
		
		& Max ($\delta$) & 0.4 & 0.7 &  & \tbcolor{} 0.3 & \tbcolor{} 0.6 &  & 1.5e+3 & 933.9 & \\
		\cline{2-4}  \cline{6-7}  \cline{9-10}
		
		& Avg ($\delta$) & 0.3 & 0.3 &  & \tbcolor{} 0.2 & \tbcolor{} 0.2 &  & 1.2e+3 & 716.0 & \\
		\cline{2-4}  \cline{6-7}  \cline{9-10}
		& Std ($\delta$) & 0.04 & 0.1 &  & \tbcolor{} 0.03 & \tbcolor{} 0.09 &  & 404.0 & 313.4 & \\			
		\hline 			
		
		\multirow{4}{*}{\textsc{Cells}} & Min($\delta$) & 13.2 & -- & \multirow{4}{*}{2404.2} & \tbcolor{} 1.3 & -- & \multirow{4}{*}{2358.5} & 2.5e+9 & -- &  \multirow{4}{*}{191050.0} \\			   
		\cline{2-4}  \cline{6-7}  \cline{9-10}
		
		& Max ($\delta$) & 155.3 & -- &  & \tbcolor{} 150.5 & -- &  & 5.1e+9 & -- & \\
		\cline{2-4}  \cline{6-7}  \cline{9-10}
		
		& Avg ($\delta$) & 63.9 & -- &  & \tbcolor{} 58.1 & -- &  & 3.1e+9 & -- & \\
		\cline{2-4}  \cline{6-7}  \cline{9-10}
		& Std ($\delta$) & \tbcolor{} 53.9 & -- &  & 57.3 & -- &  & 8.3e+8 & -- & \\
		\hline 			
		
		\multirow{4}{*}{\textsc{Engine}} & Min($\delta$) & 2.2e+4 & -- & \multirow{4}{*}{50.6} & 3.2e+4 & -- & \multirow{4}{*}{47.9} & \tbcolor{} 2.8e+3 & -- &  \multirow{4}{*}{197.6} \\			   
		\cline{2-4}  \cline{6-7}  \cline{9-10}
		
		& Max ($\delta$) & 1.7e+5 & -- &  & \tbcolor{} 5.4e+4 & -- &  & 4.2e+14 & -- & \\
		\cline{2-4}  \cline{6-7}  \cline{9-10}
		
		& Avg ($\delta$) & 6.6e+4 & -- &  & \tbcolor{} 4.2e+4 & -- &  & 1.3e+13 & -- & \\
		\cline{2-4}  \cline{6-7}  \cline{9-10}
		& Std ($\delta$) &  4.8e+4 & -- &  & \tbcolor{} 5.5e+3 & -- &  & 7.4e+13 & -- & \\
		\hline 			
	
	\end{tabular}

\end{table*}
\cref{tab:result_compare} shows the summary of the results.
In columns $\delta_{O_1}$ and $\delta_{O_2}$, we observe that for all the benchmarks, the HAs learned by our algorithm (both ``W/o Type'' and ``Type'') achieved higher accuracy in terms of Avg($\delta$).
This is because of the adequate handling of the input variables and the inference of the resets at transitions.
Moreover, type annotation improves model accuracy, as shown in benchmarks \textsc{Tanks}, \textsc{Osci}, \textsc{Cells}, and \textsc{Engine}. However, in \textsc{Ball}, both methods perform equally.

We also observe that for the HAs learned by our learning algorithm,
the maximum DTW distance Max($\delta$) tends to be close to the minimum DTW distance Min($\delta$).
This indicates that trajectories generated by our learned model do not have a high deviation from the trajectories generated by the original model.
We discuss the detail later in this section. 
In contrast, in the POSEHAD algorithm, they tend to have a high difference between Min($\delta$) and Max($\delta$).
We also observe that for the HAs learned by the POSEHAD algorithm, the standard deviation Std($\delta$) is much larger than that learned from ours.
This suggests that our learning algorithm is better at generalization.
Moreover, our algorithm takes much less time than POSEHAD.\@ For instance, in the \textsc{Cells} benchmark, our algorithm takes less than one hour, whereas POSEHAD takes more than 53 hours. 

\subsubsection{Discussion for each benchmark}
\begin{wrapfigure}{R}{0pt}
	\includegraphics[width=0.33\linewidth]{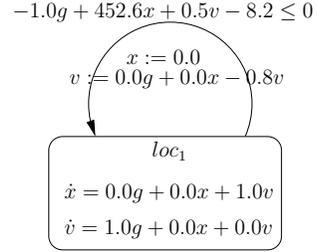}
	\caption{The HA learned by our algorithm with type annotation on \BouncingBall{}}%
	\label{fig:bballlearned}
\end{wrapfigure}
\begin{figure}
	\begin{subfigure}[t]{.3\linewidth}
		\centering
		\includegraphics[width=1.0\linewidth]{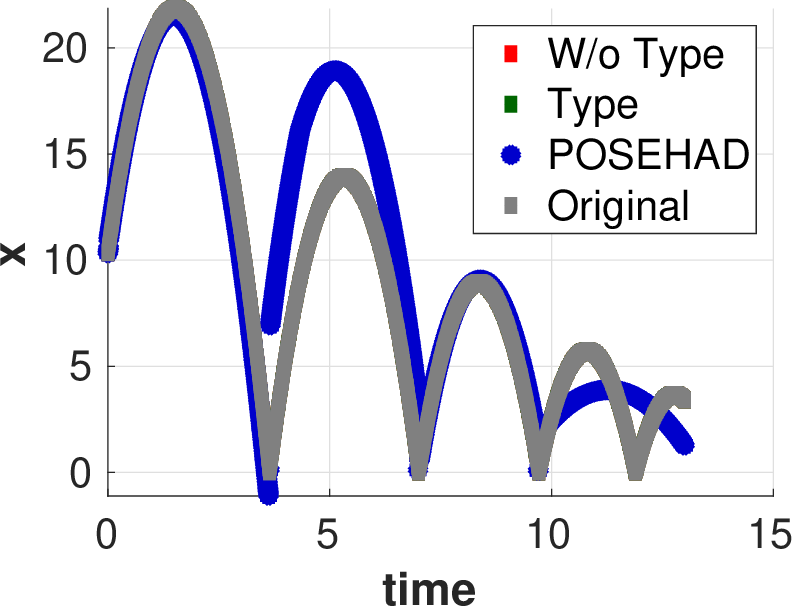}
		\caption{Position of the ball}%
		\label{fig:plotBball_position}
	\end{subfigure}	
	\hfill	
	\begin{subfigure}[t]{.3\linewidth}
		\centering
		\includegraphics[width=1.0\linewidth]{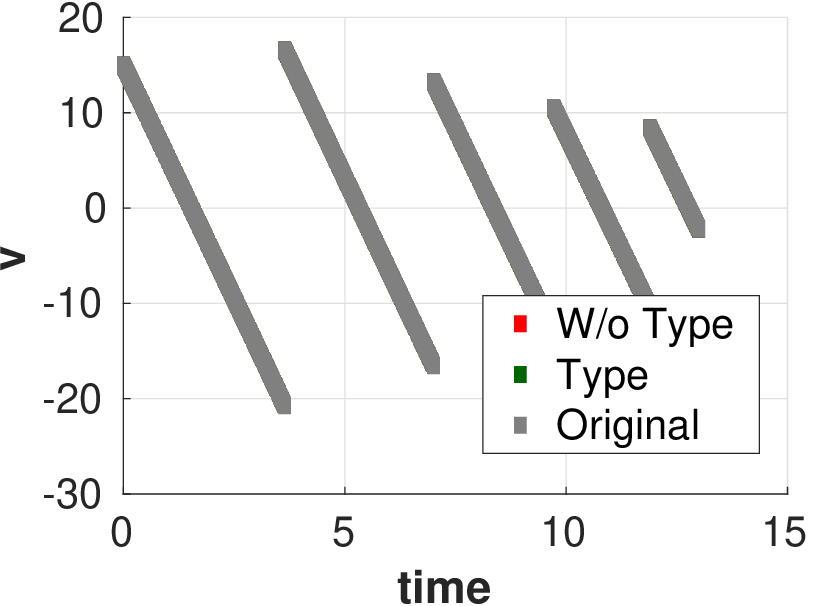}
		\caption{Velocity of the ball}%
		\label{fig:plotBball_velocity}
	\end{subfigure}
	\hfill	
	\begin{subfigure}[t]{.32\linewidth}
		\centering
		\includegraphics[width=1.0\linewidth]{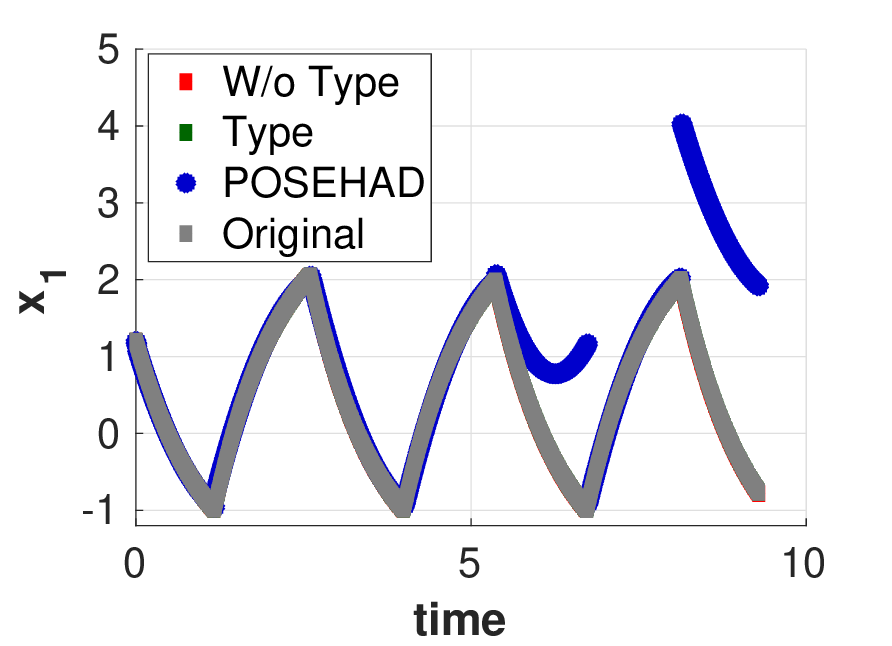}
		\caption{Liquid level $x_1$ of tank 1}%
		\label{fig:plotTwoTanksVar1}
	\end{subfigure}
	\hfill
	\begin{subfigure}[t]{.32\linewidth}
		\centering
		\includegraphics[width=1.0\linewidth]{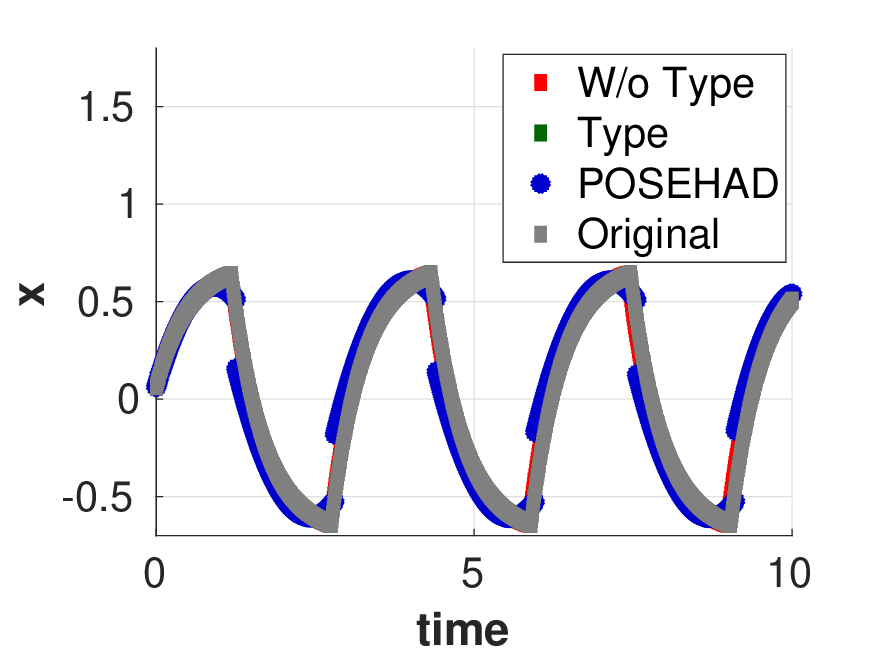}
		\caption{Position of $x$}%
		\label{fig:plotOscillatorVar1}
	\end{subfigure}
	\hfill
	\begin{subfigure}[t]{.32\linewidth}
		\centering
		\includegraphics[width=1.0\linewidth]{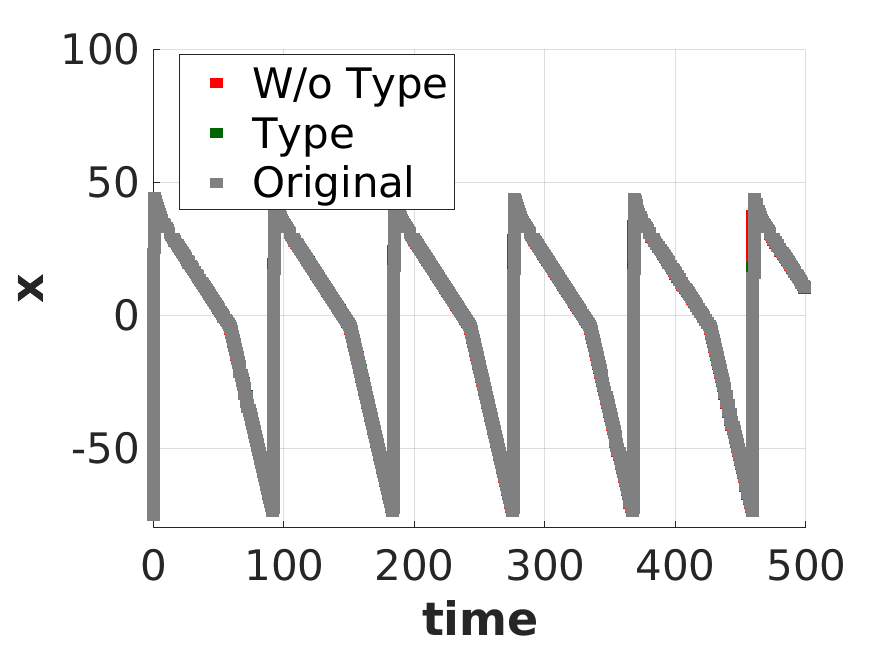}
		\caption{Voltage of the cell}%
		\label{fig:plotExcitablecells}
	\end{subfigure}
	\hfill
	\begin{subfigure}[t]{.32\linewidth}
		\centering
		\includegraphics[width=1.0\linewidth]{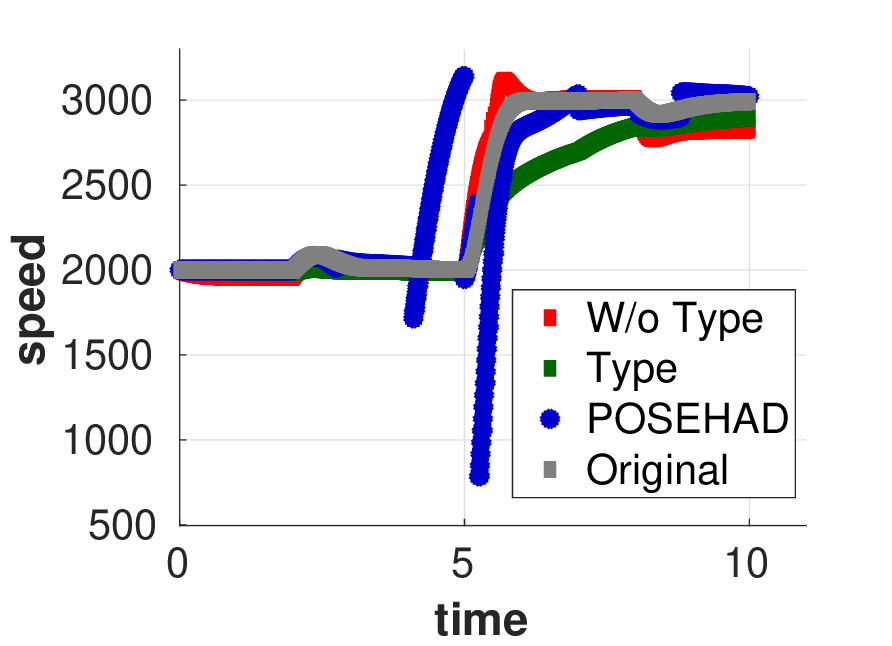}
		\caption{Engine speed}
		\label{fig:enginetiming_all_plots}
	\end{subfigure}
	\caption{Trajectories on (a-b) \textsc{Ball} (c) \textsc{Tanks} (d) \textsc{Osci} (e) \textsc{Cells} and (f) \textsc{Engine}}
	\label{figure:excitablecellsLearnedModel_illustration}
\end{figure}

\cref{fig:bballlearned} shows the learned HA for \BouncingBall{} produced by our algorithm with a type annotation.
We observe that the ODE is precisely learned.
Although the guard is far from the expected condition $x \leq 0$, it is close to the expected condition given the range of the state variables; for instance when we have $v \approx -20.55$ and $g \approx -9.8$, the condition is about $x \leq 0.020$, which is reasonably close to $x \leq 0$.
Furthermore, our algorithm accurately inferred the assignment of $v$ as $v \Coloneqq -0.8 v$. 
In \cref{fig:plotBball_position,fig:plotBball_velocity}, we show plots of the trajectories obtained from the HAs learned by our algorithm (with and without type annotation), the output trajectory predicted by POSEHAD, and the trajectory obtained from the original model.
In \cref{fig:plotBball_velocity}, we did not include the predicted trajectory by POSEHAD due to its high error.
We observe that the trajectories obtained from our learned models coincide with the original benchmark trajectory, while the trajectory predicted by POSEHAD does not. 

\begin{figure}[bp]
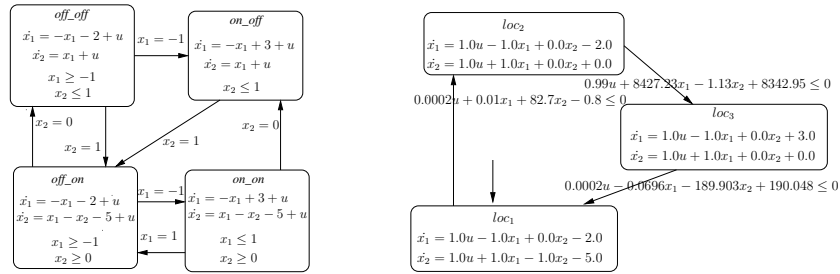

	\begin{subfigure}[t]{.33\linewidth}
		\centering
		\includegraphics[width=1.0\linewidth]{./figures/Twotanks}
		\caption{The original HA}%
		\label{fig:twotanks}
	\end{subfigure}
	\hfill	
	\begin{subfigure}[t]{.67\linewidth}
		\centering
		\includegraphics[width=0.7\linewidth]{./figures/tanks_learned}
		\caption{The HA learned by our algorithm with type annotation}%
		\label{fig:tankslearned}
	\end{subfigure}
	\caption{HAs on \TwoTankSystem{} benchmark}%
	\label{figure:TwoTanksModels}
\end{figure}	

\cref{fig:tankslearned} 
shows the HA learned by our algorithm with type annotation on the \textsc{Tanks} benchmark. 
Since the initial value, $x_2=1$, is satisfied by the guard at the initial location, the system takes an instant transition to location \textit{off\_on} (see \cref{fig:twotanks}).
Therefore, all trajectories contain data starting from this location, and our algorithm identifies this to be the initial location.
Moreover, the trajectories given to the learning algorithm do not include data visiting the location \textit{on\_on}, and this mode is not present in the learned model.
We observe that the ODEs are exactly learned, and the guards are close to the original model.
In \cref{fig:plotTwoTanksVar1}, we show a plot of the trajectories obtained from the HAs learned by our algorithm (with and without type annotation), the output trajectory predicted by POSEHAD, and the trajectory obtained from the original model.
The models learned by our algorithm produced trajectories close to the original model,
while several parts predicted by POSEHAD are far from the original one.

For the \textsc{Engine} model, due to the system's complexity, our algorithm produced HAs at most with 37 locations and 137 transitions. 
In \cref{fig:enginetiming_all_plots}, we show a plot of the trajectories obtained from the HAs learned by our algorithm (with and without type annotation), the output trajectory predicted by POSEHAD, and the trajectory obtained from the original model.
The models learned by our algorithm produced trajectories uniformly close to the original model,
 while several parts predicted by POSEHAD are far from the original one.
Similar observations on accuracy can be drawn from \cref{fig:plotOscillatorVar1,fig:plotExcitablecells} on \textsc{Osci} and \textsc{Cells} benchmarks, respectively.

\begin{figure}[tbp]
	\begin{subfigure}[t]{.5\linewidth}
		\centering
		\includegraphics[width=0.7\linewidth]{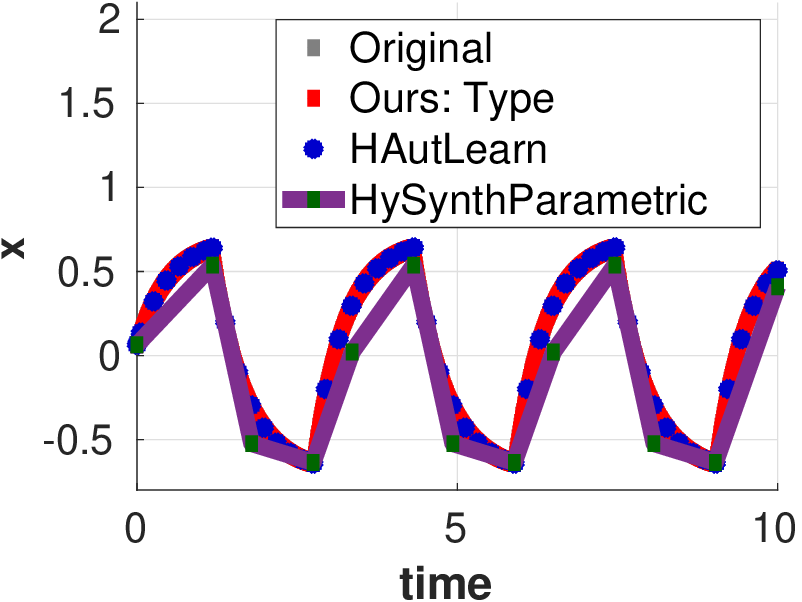}
		\caption{Variable $x$}%
		\label{fig:Compare_var_x}
	\end{subfigure}
	\hfill	
	\begin{subfigure}[t]{.5\linewidth}
		\centering
		\includegraphics[width=0.7\linewidth]{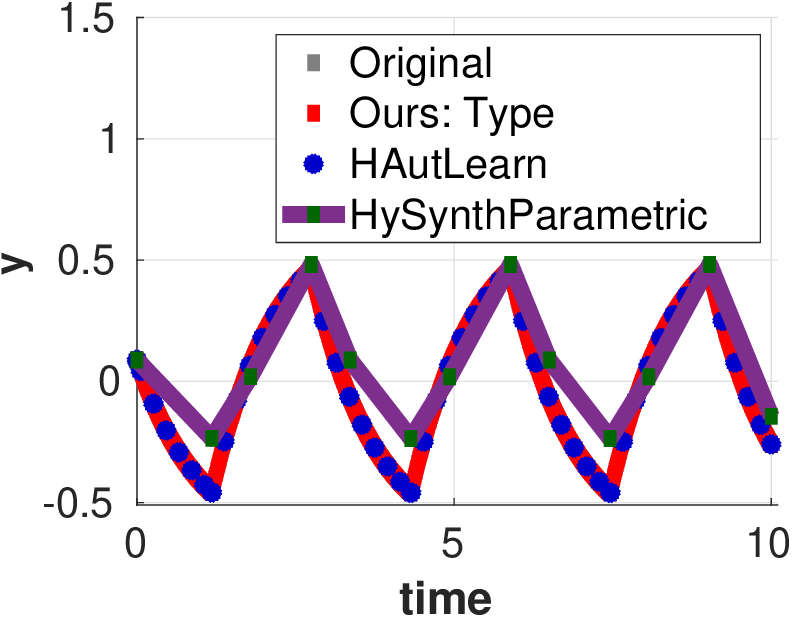}
		\caption{Variable $y$}%
		\label{fig:Compare_var_y}
	\end{subfigure}
	\begin{subfigure}[t]{.5\linewidth}
		\centering
		\includegraphics[width=0.7\linewidth]{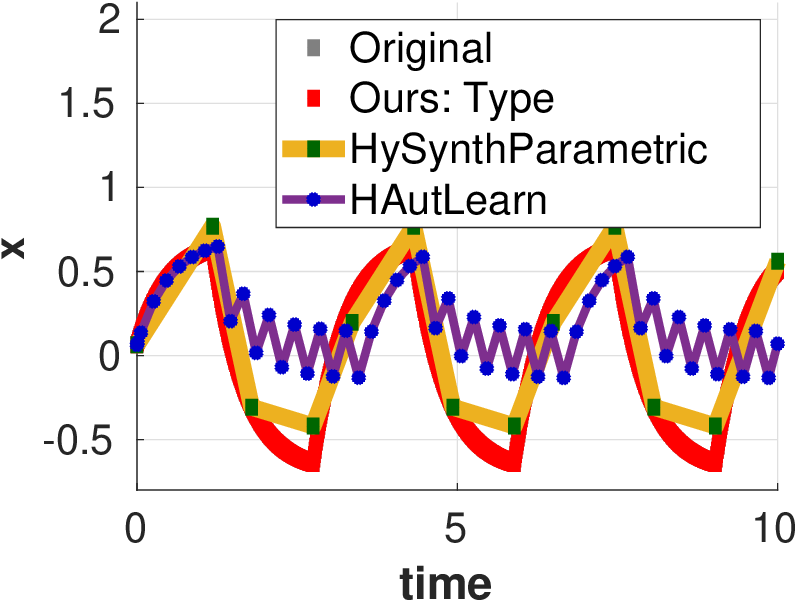}
		\caption{Variable $x$}%
		\label{fig:Compare_Osci64_var_x}
	\end{subfigure}
	\hfill	
	\begin{subfigure}[t]{.5\linewidth}
		\centering
		\includegraphics[width=0.7\linewidth]{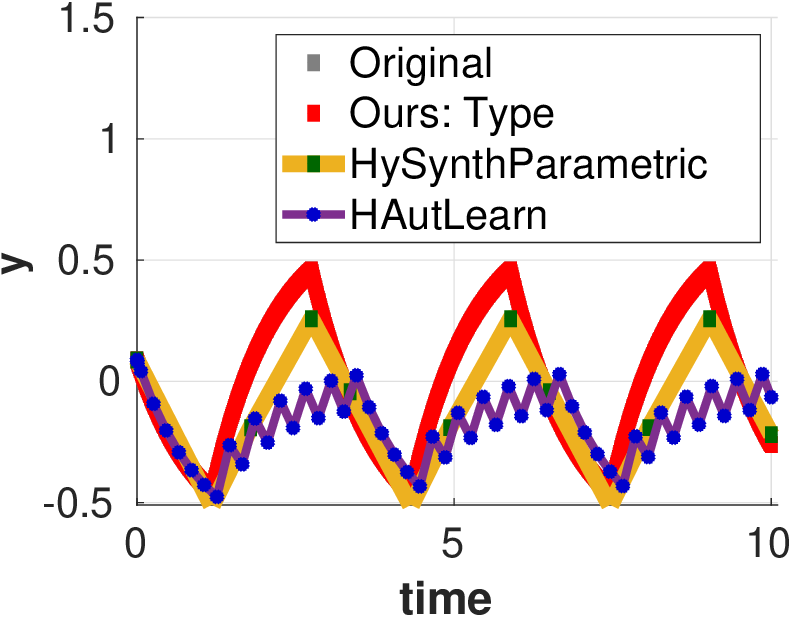}
		\caption{Variable $y$}%
		\label{fig:Compare_Osci64_var_y}
	\end{subfigure}	
	
	\caption{Trajectories obtained from the learned HAs on the \SwitchedOscillator{} benchmark by the three tools. (\protect\subref{fig:Compare_var_x}) and (\protect\subref{fig:Compare_var_y}) are trained using five trajectories, while (\protect\subref{fig:Compare_Osci64_var_x}) and (\protect\subref{fig:Compare_Osci64_var_y}) show models trained using 64 trajectories.}%
	\label{figure:toolCompare}
      \end{figure}

\subsection{Comparison with other methods}
\label{sec:comparisonWithOtherMethods}

We compared our proposed approach with other state-of-the-art methods: HAutLearn~\cite{DBLP:journals/tcps/YangBKJ22} and HySynthParametric~\cite{DBLP:conf/atva/SotoHS22}.
We conducted this experiment using only \SwitchedOscillator, which does not take an input signal, since HAutLearn and HySynthParametric do not support a model taking inputs; see \cref{table:related_work:comparison}.

The result is shown in \cref{figure:toolCompare}.
\cref{fig:Compare_var_x,fig:Compare_var_y} (resp., \cref{fig:Compare_Osci64_var_x,fig:Compare_Osci64_var_y}) show the plots obtained by the learned models trained with five (resp., 64) trajectories.
The training time with five trajectories was as follows: 60.9 seconds for HAutLearn; 15.6 seconds for HySynthParametric; 2.2 seconds for ours.
The training time with 64 trajectories was as follows: 1442.3 seconds for HAutLearn; 1483.6 seconds for HySynthParametric; 24.9 seconds for ours.

For the experiment with five training trajectories, the HA learned by HAutLearn is as precise as our method.
However, for the experiment with 64 training trajectories, we observed that the switching guard in the HA learned by HAutLearn allowed the model to take an early jump from the second jump onwards, thus generating plots that do not coincide with the original trajectories.
We can observe that the HA learned by HySynthParametric is less precise than ours.

The performance of HAutLearn was largely affected by the values of multiple parameters.
The plots in \cref{figure:toolCompare} is obtained by tuning parameters through trials and errors.

\section{Conclusion}\label{conclusion}

This paper presents an algorithm to learn an HA with polynomial ODEs from input--output trajectories.
We identify the locations by segmenting the given trajectories, clustering the segments, and inferring ODEs.
We learn transition guards using SVM with a polynomial kernel and assignment functions using linear regression.
Our experimental evaluation suggests that our algorithm produces more accurate HAs than the state-of-the-art algorithms.
Moreover, we extended the inference of assignments with type annotations to utilize prior knowledge of a user.
In future work, we plan to utilize our learned HA model to perform black-box checking~\cite{DBLP:conf/forte/PeledVY99,DBLP:conf/hybrid/Waga20,DBLP:conf/rv/ShijuboWS21} for falsification, model-bounded monitoring of hybrid systems~\cite{10.1145/3529095}, and controller synthesis~\cite{DBLP:conf/adhs/SaoudJZG18,ISCNR2016}.

%
%
\begin{FinalVersionBlock}
\subsubsection{Acknowledgements}
We are grateful to the anonymous reviewers for their valuable comments.
This work was partially supported by
JST CREST Grant No.\ JPMJCR2012,
JST PRESTO Grant No.\ JPMJPR22CA,
JST ACT-X Grant No.\ JPMJAX200U, and
JSPS KAKENHI Grant No.\ 22K17873 \& 19H04084.
\end{FinalVersionBlock}

%
%
%
\bibliographystyle{splncs04}
\bibliography{mybibliography}

\begin{thebibliography}{10}
\providecommand{\url}[1]{\texttt{#1}}
\providecommand{\urlprefix}{URL }
\providecommand{\doi}[1]{https://doi.org/#1}

\bibitem{MathWorksEngineTiming}
{MathWorks: Engine Timing Model with Closed Loop Control}.
  \url{https://in.mathworks.com/help/simulink/slref/engine-timing-model-with-closed-loop-control.html},
  accessed: 2022-12-29

\bibitem{MathWorksBouncingBall}
{MathWorks: Simulation of Bouncing Ball}.
  \url{https://in.mathworks.com/help/simulink/slref/simulation-of-a-bouncing-ball.html},
  accessed: 2022-12-29

\bibitem{AlurCHHHNOSY95}
Alur, R., Courcoubetis, C., Halbwachs, N., Henzinger, T.A., Ho, P.H., Nicollin,
  X., Olivero, A., Sifakis, J., Yovine, S.: The algorithmic analysis of hybrid
  systems. Theoretical Computer Science  \textbf{138}(1),  3--34 (1995)

\bibitem{bellman1959adaptive}
Bellman, R., Kalaba, R.: On adaptive control processes. IRE Transactions on
  Automatic Control  \textbf{4}(2), ~1--9 (1959)

\bibitem{DBLP:conf/sfm/BortolussiP08}
Bortolussi, L., Policriti, A.: Hybrid systems and biology. In: Bernardo, M.,
  Degano, P., Zavattaro, G. (eds.) Formal Methods for Computational Systems
  Biology, 8th International School on Formal Methods for the Design of
  Computer, Communication, and Software Systems, {SFM} 2008, Bertinoro, Italy,
  June 2-7, 2008, Advanced Lectures. Lecture Notes in Computer Science,
  vol.~5016, pp. 424--448. Springer (2008).
  \doi{10.1007/978-3-540-68894-5\_12},
  \url{https://doi.org/10.1007/978-3-540-68894-5\_12}

\bibitem{butcher2016numerical}
Butcher, J.C.: Numerical methods for ordinary differential equations. John
  Wiley \& Sons (2016)

\bibitem{ISCNR2016}
Filippidis, I., Dathathri, S., Livingston, S.C., Ozay, N., Murray, R.M.:
  Control design for hybrid systems with tulip: The temporal logic planning
  toolbox. 2016 IEEE Conference on Control Applications (CCA)  (Sep 2016).
  \doi{10.1109/cca.2016.7587949},
  \url{http://dx.doi.org/10.1109/CCA.2016.7587949}

\bibitem{FLGDCRLRGDM11}
Frehse, G., Le~Guernic, C., Donz\'e, A., Cotton, S., Ray, R., Lebeltel, O.,
  Ripado, R., Girard, A., Dang, T., Maler, O.: {SpaceEx: Scalable Verification
  of Hybrid Systems}. In: Ganesh~Gopalakrishnan, S.Q. (ed.) Proc. 23rd
  International Conference on Computer Aided Verification (CAV). pp. 379--395.
  LNCS, Springer (2011), \url{http://spaceex.imag.fr}

\bibitem{GPV2012}
Garulli, A., Paoletti, S., Vicino, A.: A survey on switched and piecewise
  affine system identification. IFAC Proceedings Volumes  \textbf{45}(16),
  344–355 (Jul 2012). \doi{10.3182/20120711-3-be-2027.00332},
  \url{http://dx.doi.org/10.3182/20120711-3-be-2027.00332}

\bibitem{DBLP:conf/hybrid/GrosuMYERS07}
Grosu, R., Mitra, S., Ye, P., Entcheva, E., Ramakrishnan, I.V., Smolka, S.A.:
  Learning cycle-linear hybrid automata for excitable cells. In: Bemporad, A.,
  Bicchi, A., Buttazzo, G.C. (eds.) Hybrid Systems: Computation and Control,
  10th International Workshop, {HSCC} 2007, Pisa, Italy, April 3-5, 2007,
  Proceedings. Lecture Notes in Computer Science, vol.~4416, pp. 245--258.
  Springer (2007). \doi{10.1007/978-3-540-71493-4\_21},
  \url{https://doi.org/10.1007/978-3-540-71493-4\_21}

\bibitem{hiskens2001stability}
Hiskens, I.A.: Stability of limit cycles in hybrid systems. In: Proceedings of
  the 34th Annual Hawaii International Conference on System Sciences. pp.
  6--pp. IEEE (2001)

\bibitem{jin2021inferring}
Jin, X., An, J., Zhan, B., Zhan, N., Zhang, M.: Inferring switched nonlinear
  dynamical systems. Formal Aspects of Computing  \textbf{33}(3),  385--406
  (2021)

\bibitem{keller2021discovery}
Keller, R.T., Du, Q.: Discovery of dynamics using linear multistep methods.
  SIAM Journal on Numerical Analysis  \textbf{59}(1),  429--455 (2021)

\bibitem{lygeros2008hybrid}
Lygeros, J., Tomlin, C., Sastry, S.: Hybrid systems: modeling, analysis and
  control. Electronic Research Laboratory, University of California, Berkeley,
  CA, Tech. Rep. UCB/ERL M  \textbf{99}, ~6 (2008)

\bibitem{DBLP:conf/forte/PeledVY99}
Peled, D.A., Vardi, M.Y., Yannakakis, M.: Black box checking. In: Wu, J.,
  Chanson, S.T., Gao, Q. (eds.) Formal Methods for Protocol Engineering and
  Distributed Systems, {FORTE} {XII} / {PSTV} XIX'99, {IFIP} {TC6} {WG6.1}
  Joint International Conference on Formal Description Techniques for
  Distributed Systems and Communication Protocols {(FORTE} {XII)} and Protocol
  Specification, Testing and Verification {(PSTV} XIX), October 5-8, 1999,
  Beijing, China. {IFIP} Conference Proceedings, vol.~156, pp. 225--240. Kluwer
  (1999)

\bibitem{10.1145/3556543}
Saberi, I., Faghih, F., Bavil, F.S.: A passive online technique for learning
  hybrid automata from input/output traces. ACM Trans. Embed. Comput. Syst.
  \textbf{22}(1) (oct 2022). \doi{10.1145/3556543},
  \url{https://doi.org/10.1145/3556543}

\bibitem{DBLP:conf/adhs/SaoudJZG18}
Saoud, A., Jagtap, P., Zamani, M., Girard, A.: Compositional abstraction-based
  synthesis for cascade discrete-time control systems. In: Abate, A., Girard,
  A., Heemels, M. (eds.) 6th {IFAC} Conference on Analysis and Design of Hybrid
  Systems, {ADHS} 2018, Oxford, UK, July 11-13, 2018. IFAC-PapersOnLine,
  vol.~51, pp. 13--18. Elsevier (2018). \doi{10.1016/j.ifacol.2018.08.003},
  \url{https://doi.org/10.1016/j.ifacol.2018.08.003}

\bibitem{senin2008dynamic}
Senin, P.: Dynamic time warping algorithm review. Information and Computer
  Science Department University of Hawaii at Manoa Honolulu, USA
  \textbf{855}(1-23), ~40 (2008)

\bibitem{DBLP:conf/rv/ShijuboWS21}
Shijubo, J., Waga, M., Suenaga, K.: Efficient black-box checking via model
  checking with strengthened specifications. In: Feng, L., Fisman, D. (eds.)
  Runtime Verification - 21st International Conference, {RV} 2021, Virtual
  Event, October 11-14, 2021, Proceedings. Lecture Notes in Computer Science,
  vol. 12974, pp. 100--120. Springer (2021).
  \doi{10.1007/978-3-030-88494-9\_6},
  \url{https://doi.org/10.1007/978-3-030-88494-9\_6}

\bibitem{DBLP:conf/hybrid/SotoH021}
Soto, M.G., Henzinger, T.A., Schilling, C.: Synthesis of hybrid automata with
  affine dynamics from time-series data. In: Bogomolov, S., Jungers, R.M.
  (eds.) {HSCC} '21: 24th {ACM} International Conference on Hybrid Systems:
  Computation and Control, Nashville, Tennessee, May 19-21, 2021. pp.
  2:1--2:11. {ACM} (2021). \doi{10.1145/3447928.3456704},
  \url{https://doi.org/10.1145/3447928.3456704}

\bibitem{DBLP:conf/atva/SotoHS22}
Soto, M.G., Henzinger, T.A., Schilling, C.: Synthesis of parametric hybrid
  automata from time series. In: Bouajjani, A., Hol{\'{\i}}k, L., Wu, Z. (eds.)
  Automated Technology for Verification and Analysis - 20th International
  Symposium, {ATVA} 2022, Virtual Event, October 25-28, 2022, Proceedings.
  Lecture Notes in Computer Science, vol. 13505, pp. 337--353. Springer (2022).
  \doi{10.1007/978-3-031-19992-9\_22},
  \url{https://doi.org/10.1007/978-3-031-19992-9\_22}

\bibitem{DBLP:conf/cav/SotoHSZ19}
Soto, M.G., Henzinger, T.A., Schilling, C., Zeleznik, L.: Membership-based
  synthesis of linear hybrid automata. In: Dillig, I., Tasiran, S. (eds.)
  Computer Aided Verification - 31st International Conference, {CAV} 2019, New
  York City, NY, USA, July 15-18, 2019, Proceedings, Part {I}. Lecture Notes in
  Computer Science, vol. 11561, pp. 297--314. Springer (2019).
  \doi{10.1007/978-3-030-25540-4\_16},
  \url{https://doi.org/10.1007/978-3-030-25540-4\_16}

\bibitem{suli2003introduction}
S{\"u}li, E., Mayers, D.F.: An introduction to numerical analysis. Cambridge
  university press (2003)

\bibitem{DBLP:conf/hybrid/Waga20}
Waga, M.: Falsification of cyber-physical systems with robustness-guided
  black-box checking. In: Ames, A.D., Seshia, S.A., Deshmukh, J. (eds.) {HSCC}
  '20: 23rd {ACM} International Conference on Hybrid Systems: Computation and
  Control, Sydney, New South Wales, Australia, April 21-24, 2020. pp.
  11:1--11:13. {ACM} (2020). \doi{10.1145/3365365.3382193},
  \url{https://doi.org/10.1145/3365365.3382193}

\bibitem{10.1145/3529095}
Waga, M., Andr\'{e}, E., Hasuo, I.: Model-bounded monitoring of hybrid systems.
  ACM Trans. Cyber-Phys. Syst.  \textbf{6}(4) (nov 2022).
  \doi{10.1145/3529095}, \url{https://doi.org/10.1145/3529095}

\bibitem{DBLP:journals/tcps/YangBKJ22}
Yang, X., Beg, O.A., Kenigsberg, M., Johnson, T.T.: A framework for
  identification and validation of affine hybrid automata from input-output
  traces. {ACM} Trans. Cyber Phys. Syst.  \textbf{6}(2),  13:1--13:24 (2022).
  \doi{10.1145/3470455}, \url{https://doi.org/10.1145/3470455}

\bibitem{ye2005efficient}
Ye, P., Entcheva, E., Grosu, R., Smolka, S.A.: Efficient modeling of excitable
  cells using hybrid automata. In: Proc. of CMSB. vol.~5, pp. 216--227 (2005)

\end{thebibliography}
\appendix
\section{Additional Detailed}

\subsection{Experiment settings}
For our experiment in \cref{tab:result_compare}, the maximum number of trajectory segments we take for each location in the ODE inference helps our algorithm improve performance compared to POSEHAD. For the benchmarks \textsc{Ball}, \textsc{Tanks}, and \textsc{Osci}, we take $50$ trajectory segments, $100$ for \textsc{Engine}, and $3$ for \textsc{Cells}, respectively.

We thank the authors for providing us with the source code of the POSEHAD algorithm. POSEHAD also uses the DTW algorithm for clustering similar segmented trajectories. However, for segmentation, they use a different off-the-shelf Python library named Rupture to detect change points in trajectories. Therefore, the threshold parameters that we use for our algorithm may not be the best for POSEHAD. So, as recommended in their paper, we perform a simple manual grid search of parameters, including the thresholds used in our approach. We fix a parameter that performs the best and keeps the implementation running without returning errors during the entire search. In the original POSEHAD implementation, pre-processing is applied to the input-output data by scaling the data values to 0 and 1. To perform a fair comparison, we skip this pre-processing in the experiment. POSEHAD learns an HA model for each output variable independently. 

\begin{figure}[tb]
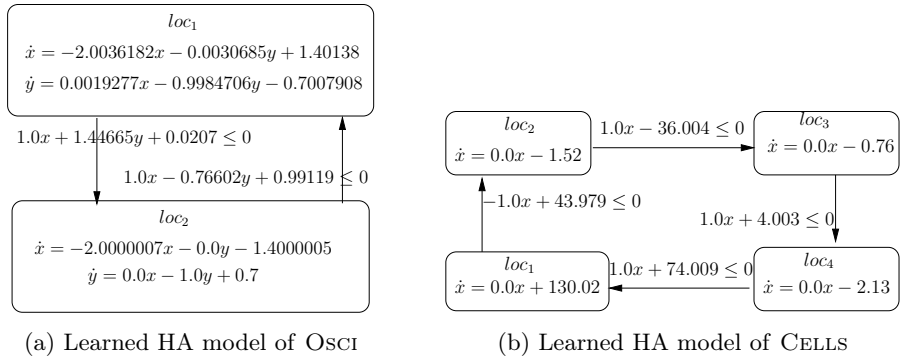


	\begin{subfigure}[t]{.45\linewidth}
		\centering
		\includegraphics[width=0.9\linewidth]{./figures/oscillator_learned}
		\caption{Learned HA model of \textsc{Osci}}
		\label{fig:oscillatorlearned}
	\end{subfigure}
	\hfill
	\begin{subfigure}[t]{.5\linewidth}
		\centering
		\includegraphics[width=1.0\linewidth]{./figures/excitableCells_learned}
		\caption{Learned HA model of \textsc{Cells}}%
		\label{fig:excitablecellslearned}
	\end{subfigure}	
	\caption{Our learned HA models using type annotation}%
	\label{figure:oscillatorModels}
\end{figure}

\subsection{Additional experimental results} \label{accuracy_results}

\subsubsection{Detailed discussion for each benchmark}

\begin{wrapfigure}{r}{0pt}
	\includegraphics[width=0.4\linewidth]{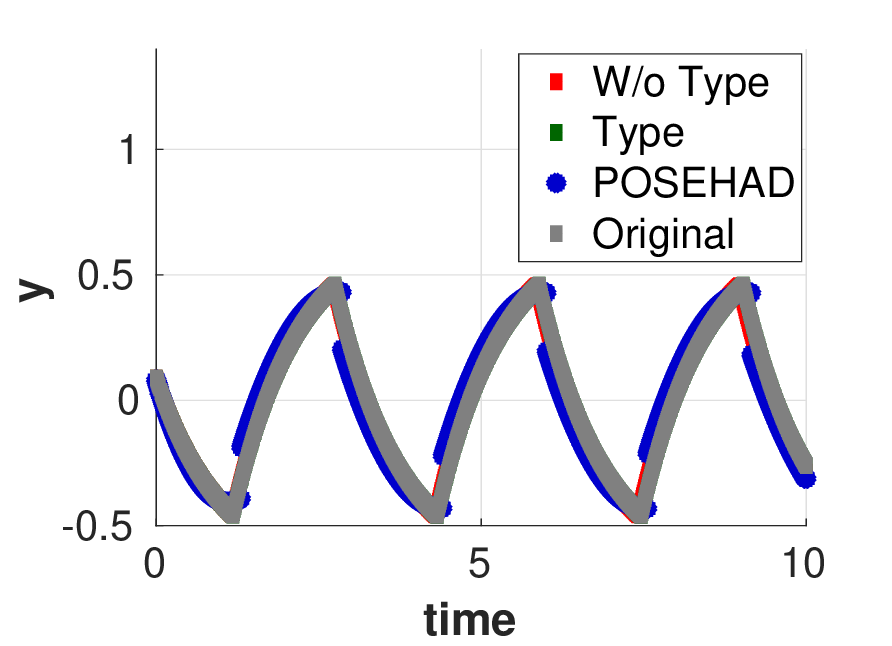}
	\caption{Trajectories of \textsc{Osci} model on variable $y$}%
	\label{fig:plotOscillatorVar2}
\end{wrapfigure}
\cref{fig:oscillatorlearned} shows our learned HA model for an \textsc{Osci} model generated using the Type annotation approach.  
Observe that in the original model, locations $loc_1$ and $loc_4$ have the same ODE, and there is no assignment logic (cause for a change point) for a discrete transition. Therefore our segmentation process considers these locations to be a single mode. Similarly, locations $loc_2$ and $loc_3$ have the same dynamics, so our approach also learns a single location for this. The ODE and the transition guards in the learned model are relatively close to the original model. 
In \cref{fig:plotOscillatorVar1,fig:plotOscillatorVar2}
, we compare output trajectories obtained by our learned models (with and without Type annotation), POSEHAD prediction, and the original benchmark. The trajectory obtained by our learned model using Type annotation overlaps precisely with the original benchmark trajectory. Without a Type annotation, the trajectory either overlaps or passes close to the original trajectory. On the other hand, several sections of the predicted trajectory by POSEHAD are either incorrectly predicted or do not overlap with the original trajectory.

\cref{fig:excitablecellslearned} shows our learned HA model for the \textsc{Cells} model produced using the Type annotation approach. We learned a four-location HA with deterministic guards for each transition. Note that the learned ODE is exact to the original model for the associated locations, and the guard conditions are close to the actual guards.
In \cref{fig:plotExcitablecells}, we show the accuracy of our learned model where trajectories generated by our learned models (with and without Type annotation) coincide with the trajectory obtained by the original benchmark. Due to high errors, we could not show the predicted trajectory by POSEHAD here in a single figure.

\ifdefined\VersionWithComments%
\setcounter{tocdepth}{1}
\listoftodos{}
\fi
\end{document}